\documentclass[pra,onecolumn,preprint,preprintnumbers]{revtex4} 

\usepackage{graphicx}
\usepackage{bm}
\usepackage{amsmath,hyperref}
\usepackage{amssymb}
\usepackage{amsfonts}
\usepackage{fancyhdr}
\usepackage{slashed}
\usepackage{latexsym,bbm}
\usepackage{amsthm}
\usepackage{float}
\usepackage{amsthm}

\usepackage{algorithm,algpseudocode}

\usepackage{enumerate}

\newcommand{\ket}[1]{\left |{#1} \right \rangle}

\usepackage{color}
\definecolor{blue}{rgb}{0,0.2,1}

\definecolor{red}{rgb}{0.9,0,0}

\newcommand{\Ord}[1]{\mathcal{O}\left( #1 \right)}
\newcommand{\tOrd}[1]{\widetilde{\mathcal{O}}\left( #1 \right)}

\theoremstyle{plain}

\newtheorem{theorem}{Theorem}

\newtheorem{lemma}{Lemma}

\newtheorem{defn}{Definition}

\def\be{\begin{eqnarray}}
\def\ee{\end{eqnarray}}

\usepackage{color}
\definecolor{Pr}{rgb}{0.4,0.3,0.9}

\begin{document}

\title{Quantum algorithm for structure learning of Markov Random Fields}
\author{Liming Zhao}\email{cqtzhao@nus.edu.sg}
\thanks{Centre for Quantum Technologies, National University of Singapore, Singapore 117543}
\author{Siyi Yang}
\thanks{Centre for Quantum Technologies, National University of Singapore, Singapore 117543}
\author{Patrick Rebentrost}\email{cqtfpr@nus.edu.sg}
\thanks{Centre for Quantum Technologies, National University of Singapore, Singapore 117543}
\date{\today }
\begin{abstract}
Markov random fields (MRFs) appear in many problems in machine learning and statistics. 
From a computational learning theory point of view, a natural problem of learning MRFs arises: given samples from an MRF from a restricted class, learn the structure of the MRF, that is the neighbors of each node of the underlying graph. 
In this work, we start at a known near-optimal classical algorithm for this learning problem and develop a modified classical algorithm. This classical algorithm  retains the run time and guarantee of the previous algorithm and enables the use of quantum subroutines. Adapting a previous quantum algorithm, the Quantum Sparsitron, 
we provide a polynomial quantum speedup in terms of  the number of variables for learning the structure of an MRF, if the MRF has bounded degree.
\end{abstract}

\maketitle

\section{Introduction}

Quantum algorithms promise speed-ups over any known classical algorithm for certain problems.  Grover provided a quantum algorithm of finding an element in a data set with a quadratic speedup over the classical search problem \cite{grover1997quantum}.  \citeauthor{durr1996quantum} presented a quantum minimum finding algorithm which is used to find the minimum element of a given data set \cite{durr1996quantum}. 
In particular settings, quantum algorithms can provide exponential speed-up, such as the HHL (\citeauthor{HHL09}) algorithm for solving linear system \cite{HHL09}, as long as a set of caveats are satisfied and a quantum state output is sufficient. 
The quantum amplitude amplification algorithm has been proposed in Ref.~\cite{brassard2002quantum}, which is based on similar techniques as Grover search. This algorithm is widely used in constructing quantum algorithms, as it can be used to estimate norms of vectors and inner products of vectors with the possibility of quadratic speedups. 
For example, given two non-negative and bounded $N$-dimensional vectors $u$ and $v$ via a quantum random access memory (QRAM) \cite{giovannetti2008quantum,arunachalam2015robustness} or efficient computation of the vector elements, amplitude estimation finds an estimate of $u\cdot v$ with error $\epsilon$ in time $\Ord{1/\epsilon}$, which is quadratically faster than the classical sampling result. 

Graphical models, which describe the dependence structure between random variables, are widely used in probability theory and machine learning. Many algorithms for learning graphical models have been developed \cite{bresler2014hardness,wu2018sparse,vuffray2019efficient,lokhov2020learning}.  A Markov random field (MRF) is a model over an undirected graph that describes a set of random variables having a Markov property. An important class of MRFs is as follows.  The MRF can be described via a distribution that is an exponential of a multi-linear polynomial of the input variables. The MRF is $r$-wise, which means that each monomial of the polynomial contains at most $r$ variables. Furthermore, an MRF with bounded degree $d$ means that the degree of each vertex of the underlying graph is at most $d$. Such MRFs are the focus of this work. 
MRFs are used in statistical physics, computer vision, machine learning and computational biology \cite{geman1986markov, clifford1990markov,diebel2005application,ma2014mrfalign}. Many algorithms for learning MRFs have been constructed \cite{bresler2013reconstruction,mckenna2019graphical,klivans2017learning,hamilton2017information}.
The problem of MRF structure learning is to discover connections between the variables, that is the presence or absence of an edge in the associated graph of the MRF. Moreover, the MRF recovery problem is to find all the connection strengths between the variables, which are the coefficients of the associated polynomial. 
A recent work Ref.~\cite{klivans2017learning} discusses such MRF learning via a multiplicative update algorithm (the ``Sparsitron"). Given $\Ord{\log n}$ samples of a binary $r$-wise MRF, the algorithm can determine the structure of the underlying graph in time $n^{\Ord{r}}$. In addition, given  $n^{\Ord{r}}$ samples, one can recover all the coefficients. 

Quantum graphical models have been studied in the last decade \cite{leifer2008quantum,srinivasan2018learning,adhikary2019learning,souissi2020forward}. In Ref.~\cite{leifer2008quantum}, the authors constructed a quantum graphical model by replacing every variable with a quantum system and applied it to quantum error correction and the simulation of quantum many-body systems.
The method of learning quantum graphical models using constrained gradient descent on the Stiefel manifold  has been explored in \cite{adhikary2019learning}. Quantum computation is promising to be more efficient for structure learning of classical graph models.  Quantum algorithms for the structure of learning classical graph models have been considered \cite{tucci2014quantum,o2015bayesian}. Learning  the structure of a Bayesian network by the quantum adiabatic algorithm has been studied in Ref.~\cite{o2015bayesian}.   Quantum algorithms for learning generalized linear models and Ising models via the Sparsitron have been studied in Ref.~\cite{rebentrost2021quantum}. It provides a polynomial speed-up in terms of the dimension of the samples over the classical algorithm.   

In this paper, we focus on the structure learning of MRFs. We first modify the classical algorithm for learning the structure of the underlying graphs of $r$-wise MRFs in Ref.~\cite{klivans2017learning} to make it easier to obtain a quantum speedup. The modification avoids the median finding part of the original algorithm and hence the construction of a quantum version of the same.
Instead of calling the Sparsitron algorithm once, we call the Sparsitron algorithm $r-1$ times. The time complexity of the modified algorithm is the same as the original one. We then construct a quantum algorithm for the same problem with one more assumption that the degree of the underlying graph is bounded.
We show that the quantum algorithm provides a polynomial speed-up in terms of the dimensionality of the MRF.

For the modified main classical MRF learning algorithm of \cite{klivans2017learning} (\textsc{MrfRecovery}) we have the following Theorem, here presented informally. 
\begin{theorem}[MRF structure learning without median finding (informal)]
For $r$-wise $n$-variable MRFs with some other parameters defined later, there exists an algorithm without median finding (Algorithm \ref{alg_classical_MRF_sparsitron}) to learn the structure of the dependency graph with high probability and $\sim \log n$ samples in run time $\Ord{n^r}$. 
\end{theorem}
The exact statement and proof are given in Theorem \ref{th:identifiable}.
This classical algorithm allows for a relatively straightforward quantum version. Our quantum algorithm  relies on a subroutine for set membership queries in quantum superposition.
The main assumption for the set membership subroutine is the existence of a quantum random access memory, see Definition \ref{defQRAM}. Given this device, for a classical vector $v$ with dimension $N$, where every element takes $m$ bits to represent, it takes time $\tOrd{N m}$ to set up the QRAM for the vector, enabling the superposition query 
\be
\ket {j} \ket{0^m} \to \ket{j} \ket {v_j}, 
\ee
for each $j \in [N]$, and each query costs $\Ord{m +\log^2 N}$ run time. 

Instead of a vector $v$, consider a set of integers. Given subset $S \subset [N]$, we can construct a QRAM for $S$. The set membership problem is to determine if an element in $[N]$ is an element in $S$, which can be done by a quantum set membership query as stated in the following Theorem. 
\begin{theorem}[Quantum set membership (informal)] \label{thmQSet_informal}
Given a set $S \subset [N]$ with size $\vert S\vert$, we can provide a unitary which performs the quantum indicator function (or set membership query)
with time  $\Ord{\log\vert S\vert\log N + \log^3\vert S \vert}$. Preprocessing requires $\tOrd{\vert S\vert \log N}$ space and time. 
\end{theorem}
The exact statement and proof are given in Theorem \ref{Quantum_indicator_function}.
Based on the modified classical MRF structure learning algorithm we construct a quantum algorithm to recover the structure of the dependency graph. The results are stated in the following theorem.

\begin{theorem}[Quantum MRF structure learning (informal)]
For $r$-wise $n$-variable MRFs with bounded degree $d$ with other parameters specified later and given quantum access to samples from the MRF, let $\mu=\min\{2^d-1,d^{r-1}\}$, there exists an quantum algorithm (Algorithm \ref{quantum_bounded-degree MRF_sparsitron}) to recover the structure of the dependency graph with high probability and $\sim \log n$ samples in run time $\tOrd{n\mu +\sqrt{\mu n^{r+1}}}$. 
\end{theorem}
The exact statement and proof are given in Theorem \ref{theorem_quantum_MRf_structure_learning}.
In the next section, we introduce notations used in this paper and the definition of MRF. In Sec.~\ref{section_classical_mrf_algorithm}, we present the modified classical algorithm for MRF structure learning.
Sec.~\ref{secqsetmrf} describes the quantum set membership queries needed for the quantum version of the MRF structure learning algorithm. Sec.~\ref{section_quantum_mrf_algorithm} describes the quantum algorithm for MRF structure learning. At last, Sec.~\ref{section_conclusion} gives the conclusion and a brief discussion on the  lower bound.
\section{Notations and preliminaries }\label{Preliminary_and_notation}

Let $\mathbb Z_+$ denote the set of positive integers and  $[N]=\{1,2,\cdots,N\}$. 
 The $1$-norm of a vector $v \in \mathbbm R^n$ is given by $\Vert v \Vert_1 =  \sum_{i=1}^n \vert v_i \vert$.
All logarithms are base $2$ and denoted by $\log$. If an algorithm costs time  $f(n) = O(n \log^k n)$ for some positive constant $k$,  we denote the run time as $\tOrd{n}$. The sigmoid function is $\sigma(z)=1/(1+e^{-z})$.

We review multi-linear polynomials. Let $Z =\left(Z_1, \cdots, Z_n\right)$ be $n$ variables from a domain, which here we take $\{-1,1\}$. 
For any subset $I \subseteq [n]$ and a function $p_I: \{-1,1\}^n \to \mathbbm R$, the subscript notation means that $p_I$ considers only the variables of $\{-1,1\}^n$ which are inside the subset $I$. An important case is the monomial function $p_I(Z) =\widehat{p}(I) Z_I$,  with the coefficient $\widehat{p}(I)$ and  the  monomial $Z_I=\prod_{k\in I}Z_k $.
A multi-linear polynomial is a sum of monomial functions, i.e., $p(Z) = \sum_{I\subseteq [n]} p_I(Z)$.   If all of the monomials of the multi-linear polynomial $p$ contain at most $r$ variables and there is at least one monomial containing exactly $r$ variables, then $r$ is  the degree of the polynomial $p$.
For a polynomial $p$, we denote $\bm p$ as the coefficient vector, where we use boldface for the coefficient vector of a polynomial.
As the $1$-norm of a vector, the $1$-norm of a polynomial $p$ is denoted as $\Vert \bm p \Vert_1=\sum_{I \subseteq [n]}\vert \widehat{p}(I)\vert$, which is the sum of the absolute value of all coefficients.
For a $u \in[n]$, consider the partial derivative 
\be \label{eqPartial}
\partial_u p(Z) := \sum_{J \subseteq [n] : u \not\in J }\widehat p \left(J \cup \{u\}\right) \prod_{j\in J}Z_j.
\ee
In this work, an important notion is that of a maximal monomial of a polynomial.
For any polynomial $p(Z)$, define a maximal monomial to be an $I\subseteq [n]$ for which $\widehat p(J) =0$ for all $J \supset I$. In words, there is no non-zero monomial that strictly contains $I$.

An undirected graph is denoted by $G\left(V, E\right)$, where $V$ is the vertex set and $E$ is the edge set. We denote $N(u)$ as the set of neighbors of a vertex $u$ of the graph $G$. The number of elements in $N(u)$ is defined as the degree of vertex $u$, which is denoted as $d_u$. We say a graph $G$ is $d$-degree bounded if $d=\max\limits_{u \in V} \{d_u\}$.  
A clique $C$ of graph $G$ is defined as a subset of vertices that are fully connected. Denote $C_r(G)$ as the set of all the cliques of $G$ with size at most $r$.  A clique is a \textit{maximal} clique if there is no other clique that contains the clique.  
A maximal clique is also called a hyperedge of graph $G$.

A Markov random field (MRF) is also known as an undirected graphical model. We can characterize a binary MRF by an $n$-vertex undirected graph $G(V, E)$ and a distribution $\mathcal D$ on $\{-1,1\}^n$. 
An MRF is an $r$-wise MRF when the number of vertices in any of the maximal cliques of $G$ is at most $r$.
The joint probability of distribution $\mathcal D$ of the $r$-wise MRF is given by an exponential family defined by the monomial functions $p_I$ as
\begin{eqnarray}\label{eqMRFdist}
\mathbbm P[Z=z] \propto \exp\left(\sum_{I \in \mathcal S}p_I(z)\right),
\end{eqnarray}
where $\mathcal S \subseteq C_r(G)$.
The function  $p(z):=\sum_{I \in \mathcal S \subseteq C_r(G)}p_I(z)$ is also called the factorization polynomial of the MRF. Note that the degree of the underlying graph and the degree of the factorization polynomial are different. For a $r$-wise MRF, we see that the degree of the factorization polynomial is $r$.  If the degree of the underlying graph is bounded by $d$, then $r\leq d-1.$  
Learning an MRF is to recover the cliques and the coefficients of the factorization polynomial.  To obtain a learning algorithm that retrieves a unique underlying graph with provable guarantees, one assumes an MRF that satisfies the following conditions.
\begin{defn}[Special class of MRFs \cite{klivans2017learning}] \label{defMRFclass}
For a $r$-wise binary MRF with $\mathcal D$ on $\{-1,1\}^n$ and $\eta,\lambda>0$, we assume that:
\begin{itemize}
\item[1.)]
The associated underlying graph $G$ and factorization Eq.~(\ref{eqMRFdist})
is $\eta$-identifiable, which is defined as: for every maximal monomial $J$ in $p(z)$, the coefficient of the monomial satisfies $|\widehat{p}(J)| \geq \eta$ and every edge in $G$ is covered by a non-zero monomial of $p$.
\item [2.)] The coefficients of the factorization polynomial are bounded as $\Vert \partial_u p\Vert_1 \leq \lambda$ for all $u\in [n]$.
\end{itemize}
\end{defn}
 
\section{Classical MRF structure learning by  Sparsitron algorithm}
\label{section_classical_mrf_algorithm}

We now describe how to recover the structure  of an $r$-wise and $\eta$-identifiable MRF using the Sparsitron algorithm. Recall that a $r$-wise MRF can be related to a $r$-order polynomial $p$. We can thus learn the structure of an MRF by learning some of the coefficients of the corresponding polynomial.
In Ref.~\cite{klivans2017learning}, the authors use the Sparsitron algorithm to learn $r$-wise MRFs. We construct a modified version of their algorithm (Algorithm $3$ in Ref.~\cite{klivans2017learning}), which avoids the use of median finding.  We present the guarantee of this algorithm and show that it exhibits the same time complexity as the previous algorithm. In the next section, we give a quantum algorithm based on our modified algorithm for bounded-degree MRF.

Let the polynomial of a $r$-wise $\eta$-identifiable MRF be $p(Z)$, as defined in Section \ref{Preliminary_and_notation}.
With the partial derivative Eq.~(\ref{eqPartial}), define the polynomial 
\be
p_u(Z) := -2 \partial_u p(Z)\label{definition_pu},
\ee
The corresponding vector form is 
\be \label{eqPuvec}
\bm{p}_u \equiv \left(\widehat{p}_u(I): I\in [n]\setminus \{u\}, \vert I \vert\leq r-1 \right) \in \mathbb R^K,
\ee
where $K=\sum_{k=1}^{r-1}  \tbinom{n-1}{k}$,  and 
using the corresponding coefficients $\widehat p (J \cup \{u\})$.  
In addition, define a vector of multi-linear monomials of degree at most $r-1$ as  
\be
X_u(Z) := \left (\prod_{i \in I} Z_i: I \subseteq [n] \setminus \{u\}, |I| \leq r-1 \right).
\ee
It is easy to see that $-2\partial_u p(Z) = \bm{p}_u \cdot X_u(Z)$.
Denote $Z_{-u}$ as all the variables except the variable $Z_u$.
We would like to learn on the space of monomials $X_u(Z_{-u})$ to obtain the elements of $\bm p_u$.
\begin{defn}\label{definition_distribution_Dprime}
Given the distribution $Z\sim \mathcal D$ on $\{-1,1\}^n$ as in Definition \ref{defMRFclass} and fixed $u\in[n]$. 
Define the distribution $\mathcal D'$ on $\{-1, 1\}^{K} \times \{0,1\}$ such that for $Z \sim \mathcal D$ we have $\left(X_u(Z),Y(Z)\right) \sim \mathcal D'$ with $Y_u(Z)=\left(1-Z_u\right)/2$.
\end{defn}
By Definition \ref{defMRFclass}, Point 2.), we have that $\Vert \bm p_u \Vert_1 \leq 2 \lambda$.
From Lemma \ref{lemma_MRF_function_unbiased},  we have
\begin{eqnarray}\label{equ_mrf}
\mathbb{ E}_{(X,Y) \sim \mathcal D'}\left[Y | X \right] = \sigma( \bm{p}_u \cdot X).
\end{eqnarray}
The MRF algorithm requires an algorithm (the Sparsitron) which probabilistically and approximately learns a polynomial $q$ which approximates the polynomial $p_u$,
such that 
\begin{eqnarray}
\mathbb{E}_{(X,Y)\sim \mathcal D'} \left[\left(\sigma( \bm p_u\cdot X  )-\sigma(\bm{q}\cdot X)\right)^2\right]<\epsilon.\label{Equ_condition_expectation_mrt}
\end{eqnarray} 
This guarantee for $\bm q$ can be translated to a guarantee for those elements of the vector $\bm q$ which belong to the maximal monomials of the polynomial $ 
p_u - q$ (cf. Section \ref{Preliminary_and_notation} for the definition of maximal monomials). 

\subsection{Maximal monomials}

The structure of an MRF can be learned by learning some coefficients of the factorization polynomial. Let $p_u:\mathbb{R}^n\rightarrow \mathbb{R}$ be a polynomial where each monomial contains at most $r-1$ variables. For $l\in [r-2]$, given the maximal monomials of $p_u$ with size strictly greater than $l$, if a monomial of $p_u$ of size exactly $l$ is not a sub monomial of any  maximal monomials with larger sizes,  it is either a maximal monomial or the coefficient is zero.    
  
We first define three types of subsets of the monomials of any polynomial in the following.
\begin{defn}\label{Subset_classical_JWF}
For $l\in [r-1]$, fix $u\in[n]$, and let any multi-linear polynomial $q$ on variables indexed by $[n]\setminus \{u\}$, where each monomial contains at most $r-1$ variables. When unclear, we use the notation $J_l(q)$, $W_l(q)$, and $F_l(q)$ to denote which polynomial we refer to. 
\begin{itemize}
    \item [(1)] 
[Maximal monomials] Define the set $J_{r-1}:=\emptyset$ and, for $l\in[r-2]$, define $J_l$ as the set of subsets $I\subseteq [n] \setminus \{u\}$ with the following conditions:
\begin{itemize}
\item[•] For all subset $I\in {J}_l$, we have   $l<\vert I\vert \leq r-1$. 
\item[•] For all subset $I\in J_l$, $I$ is a maximal monomial of $q$.
\end{itemize}
\item [(2)] [Maximal monomials or contained in maximal monomial] Define the set $W_{r-1}:=\emptyset$ and, for $l\in[r-2]$, define $W_l$ as the set of the subsets $I\subseteq[n]\setminus \{u\}$ with the following conditions:
\begin{itemize}
\item[•] For all $I\in W_l$, we have   $l<\vert I\vert \leq r-1$. 
\item[•] For all $I\in W_l$,  $\exists  A \in J_l$ such that $ I \subseteq  A$. 
\end{itemize}
\item[(3)] [Candidates for maximal monomials] Define the set $F_l$ as the set of subsets $I\subseteq [n]\setminus \{u\}$ with the following conditions:
\begin{itemize}
\item[•] For all $I \in F_l $,  $\vert I\vert = l$.
\item[•] For all $I \in F_l $, we have   $ \forall  A \in W_l, I  \nsubseteq  A $.
\end{itemize}
\end{itemize}
\end{defn}

The interpretation of these sets is as follows. The set $J_l$ are the maximal monomials we found so far by exploring the monomials of degree between $l+1$ to $r-1$. 
The set $W_l$ describes all the subsets of size $l+1$ to $r-1$ of these maximal monomials, so these sets themselves and any subsets of them are not candidates for new maximal monomials. 
Finally, the set $F_l$ describes the set of candidates which potentially can be new maximal monomials with degree $l$. 
The interpretation as candidates follows from the next Lemma \ref{lemma_two_cases_of_pu}.
For this set, we have to use further tests to determine if they are indeed maximal monomials of the polynomial $p_u$. 
For example, let $n=4$, $r=3$, $u=1$, and $p_u=Z_3+1.5Z_4+2Z_2Z_3$, when $l=1$, as Definition \ref{Subset_classical_JWF}, we have $J_1(p_u)=\{ \{2,3\}\}$, $W_1(p_u)=\{ \{2,3\}\}$ and $F_1(p_u)=\{ \{4\}\}$.

\begin{lemma}\label{lemma_two_cases_of_pu}
Let there be given a multi-linear polynomial $p_{u}$ containing monomial of which the degree at most $r-1$.
Let $l\in[r-1]$, and define the sets $J_l(p_u)$, $W_l(p_u)$, and $F_l(p_u)$ from Definition \ref{Subset_classical_JWF} in relation to $p_u$. 
Define the polynomial 
\be 
p_{u,l} &:=& \sum_{|I|\leq l}\widehat{p}_u(I)Z_I +\sum_{I\in W_l(p_u)}\widehat{p}_u(I)Z_I.
\end{eqnarray}
Then, for each $I\in F_l$, it holds that (1) $I$ is either a maximal monomial of $p_{u,l}$  or $\widehat{p}_{u,l}(I)=0$.
Let there be given another multi-linear polynomial $q_l$ with monomial degree at most $r-1$, which is
\be \label{eq:ql}
 {q_l} &:=& \sum_{|I|\leq l}\widehat{q}_l(I)Z_I +\sum_{I\in W_l(p_u)}\widehat{q}_l(I)Z_I.
 \ee
Then, for each $I\in F_l$, it holds that (2) $I$ is either a maximal monomial of $p_{u,l}-q_l$ or $\widehat p_{u,l}(I)-\widehat q_l(I)=0$.
\end{lemma}
\begin{proof}
We first consider the case that $l=r-1$.
By definition, $W_{r-1} = \emptyset$ and $F_{r-1}$ contains all subsets of $[n]\setminus \{u\}$ with size $r-1$.
From the hypothesis that that $p_u$ and $q$ have monomial degree of at most $r-1$, it follows immediately that for $I \in F_{r-1}$ the conclusion (1) is true.
Since the difference between the two polynomials does not increase the monomial degree, conclusion (2) is also true. 

Now turn to  $l<r-1$. For each $l\in[r-2]$, by Definition \ref{Subset_classical_JWF},
$W_l$ contains all monomials with size greater than $l$ and which are either maximal monomials or contained in a maximal monomial. 
The set $F_l$ contains monomials with size $l$ and which are not a subset of any element in $W_l$ (defined via $p_u$), so the conclusion (1) is true. 
We can see that a subset in $F_l$ is not contained in any monomials with size greater than $l$ also in $q_l$, by definition of $q_l$ in Eq.~(\ref{eq:ql}). Hence, conclusion (2) is also true.
\end{proof}

\subsection{Maximal monomial approximation by the Sparsitron algorithm}

It is straightforward to see that $\mathcal D'$ with Eq.~(\ref{equ_mrf}) defined in the beginning of this Section \ref{section_classical_mrf_algorithm} satisfies the hypothesis on the distribution of the \textsc{Sparsitron} (see Theorem \ref{theorem_sparsitron}).
The variables $Z$ without the variable $Z_u$ become the ``features" in the Sparsitron, while $Z_u$ transforms to the ``label" for the \textsc{Sparsitron}.
We satisfy the other hypothesis of the Theorem \ref{theorem_sparsitron} because $\Vert \bm p_u \Vert_1 \leq 2 \lambda$. 
Then by the algorithm of Theorem \ref{theorem_sparsitron} we can obtain a polynomial $q$ according to Eq.~(\ref{Equ_condition_expectation_mrt}).
The \textsc{Sparsitron} algorithm is shown in Algorithm \ref{algosparsitron} in Appendix \ref{app_classical_sparsitron}. It is a modified version of the celebrated Hedge and Adaboost algorithms by Freund and Schapire \cite{freund1997decision}.
The \textsc{Sparsitron} algorithm can be used to learn sparse Generalized Linear Models and Ising models. It uses a multiplicative weight update rule in contrast to other algorithms which use additive update rules. The guarantee of the \textsc{Sparsitron} is shown in the following theorem.  
\begin{theorem}[\textsc{Sparsitron} \cite{klivans2017learning}]\label{theorem_sparsitron}
Let $\mathcal D$ be a distribution on $(X,Y) \in [-1,1]^n \times \{0,1\}$, for which $\mathbb{ E}[Y\vert X=x]=\sigma(w\cdot x)$ for a 
non-decreasing $1$-Lipschitz function $\sigma: \mathbbm R\to [0,1]$ and $w \in \mathbbm R^n$. Suppose that $\Vert w \Vert_1\leq \lambda $ for a known $\lambda \geq 0$. Then, there exists an algorithm that for all $\varepsilon,  \rho \in(0,1)$ given $T\in \Ord{\lambda^2(\log(n/\rho\epsilon))/\epsilon^2}$ independent samples from $\mathcal D$, produces a vector $v\in \mathbbm R^n$ such that with probability at least $1-\rho$,
 $$ \mathbb{ E}\left[ \left(\sigma (v\cdot X)-\sigma (w\cdot X) \right)^2\right] \leq \epsilon.$$ 
 The run-time of the algorithm is $\Ord{nT}.$
\end{theorem}
Note that we slightly abuse the notation for the function $\sigma$: in the theorem, it is any $1$-Lipschitz function, while in the remainder of this work it is the sigmoid function.
The Theorem \ref{theorem_sparsitron} is stated for vectors $w,v \in \mathbbm R^n$ containing positive and negative elements. 
On the other hand, Algorithm \ref{algosparsitron} constructs a positive weight vector based on multiplicative updates. From the algorithm output, the theorem statement can be obtained via a simple trick \cite{klivans2017learning}, considering an enlarged learning problem. For every $x\in [-1,1]^n$ use the map $\widetilde{x}=(x,-x,0)$ to transform the input space $(x,y)$ to $(\widetilde{x},y)$. The Sparsitron Algorithm \ref{algosparsitron} applied to the enlarged learning problem returns a non-negative vector. Let this vector be denoted by  $\widetilde{v}=(v_1, v_2, v_3)$, with $v_3 \in\mathbbm R$. For this vector $\Vert \tilde v \Vert_1 = \lambda$. This vector can be mapped to the vector $v=v_1-v_2$ which is the estimation of the original vector $w$, with $\Vert v\Vert_1 \leq \lambda$. Note the identity $\widetilde{v}\cdot \widetilde{x} = v_1 \cdot x - v_2 \cdot x = v \cdot x$.

Assuming we have run the Sparsitron we obtain the guarantee Eq.~(\ref{Equ_condition_expectation_mrt}) for the resulting polynomial $q$.
We show which choice of $\epsilon$ in Eq.~(\ref{Equ_condition_expectation_mrt}) allows to find  all the maximal monomials of $p_u$ from the polynomial $q$. 
From Lemma \ref{lemma_6.3} we can show that for a maximal monomial $I$ of $p_u-q$, the coefficient $\widehat p_u (I)$ can be estimated by $\widehat q(I)$ with bounded error. 
\begin{lemma}\label{lemma_threshold_eta0}
Let $\cal{D}$ be the distribution of an $r$-wise MRF on $\{1,-1\}^n$ that is $\eta$-identifiable. Fix $u\in[n]$ which specifies the unknown polynomial $p_u$ defined in Eq.(\ref{definition_pu}). With $\lambda,\epsilon>0$, let $\Vert p_u\Vert_1 \leq 2\lambda$ and let a polynomial $q$ satisfy $\mathbb{E}_{Z\sim \cal{D}}\left[\left(\sigma( p_u(Z)) -\sigma(q(Z))\right)^2\right]\leq \epsilon$, where
$\epsilon < e^{-6 -4\lambda -2\lambda(r-1)} \eta^2/(2^{r+1})$. 
For any subset $I \subseteq [n]\setminus \{u\}$ that is a maximal monomial of $p_u-q$, it holds that
\be
\vert \widehat{ p}_u(I)- \widehat q(I)\vert \leq \frac{\eta}{2}.
\ee
\end{lemma}
\begin{proof}
Let $\delta = e^{-2\lambda}/2$. Note that the $\epsilon$ from the hypothesis is smaller than the hypothesis on $\epsilon$ of Lemma \ref{lemma_6.3}, $\epsilon <  e^{-2\Vert p_u \Vert_1 -6} \delta^{r-1} \eta^2/4$. Then Lemma \ref{lemma_6.3} implies that for a maximal monomial $I$ of the polynomial $p_u-q$,  the corresponding coefficients satisfy  $\vert  \widehat p_u(I) -\widehat{q}(I)\vert \leq  e^{\Vert p_u \Vert_1 +3}\sqrt{\epsilon / \delta^{\vert I\vert}}$.    
 Since  $\epsilon < e^{-2\Vert p_u \Vert_1 -6} \delta^{r-1} \eta^2/4$, we have
 \begin{eqnarray}
\vert \widehat{ p}_u(I)- \widehat q(I)\vert  \leq \frac{\eta}{2}\sqrt{\frac{ \delta^{r-1}}{ \delta^{|I|} }}  \leq \frac{\eta}{2}, \label{eq_coefficient_difference}
\end{eqnarray}
where the last inequality follows from  $\delta <1$, and $\vert I \vert \leq r-1$.
\end{proof}

By the definition of $\eta$-identifiable MRF, the absolute value of  coefficients of the factorization polynomial $p$ are either no less than $\eta$ or equal to $0$ for all maximal monomials of $p$.  Then the absolute value of coefficients of all maximal monomial of $p_u$ is no less than $2\eta$  as $p_u = -2 \partial_u p$.

\begin{lemma}\label{lemma_threshold_eta1}
Let $l\in [r-1]$ and $\eta>0$. Let there be given an $\eta$-identifiable polynomial $p$ with $p_u$ according to Definition \ref{Subset_classical_JWF} for $u\in [n]$, 
and 
the sets $J_l$, $W_l$, and $F_l$ defined in relation to $p_u$. 
Let $p_{u,l}$ and $q_l$ as in Lemma \ref{lemma_two_cases_of_pu}.  
In addition, let it hold that 
for any subset $I' \subseteq [n]\setminus \{u\}$ that is a maximal monomial of $p_{u,l}-q_l$ we have
\be \label{eq:pu_q_guarantee}
\vert \widehat{ p}_{u}(I')- \widehat q_l(I')\vert \leq \frac{\eta}{2}.
\ee
Then, for any subset $I\in F_l$ it holds that: 
\begin{itemize}
\item[1)] If  $\vert \widehat{q}_l(I) \vert \geq \eta $,   $I$ is a maximal monomial of $p_u$ which implies $ \vert \widehat{p}_u(I) \vert \geq 2\eta $ , 
\item[2)] If  $\vert \widehat{q}_l(I) \vert < \eta ,$ then we have $\vert \widehat{p}_u(I) \vert = 0$
\end{itemize}
\end{lemma}

\begin{proof}
Take $I \subseteq [n]\setminus \{u\}$ such that it is a maximal monomial of $p_{u,l}-q_l$.
For the case $\vert \widehat{q_l}(I) \vert \geq\eta$,  we can show the lower bound
\be
\vert\widehat{p}_u(I) \vert \geq \vert\widehat{q_l}(I) \vert  - \vert  \widehat q_l(I)-\widehat{ p}_u(I) \vert
\geq \eta - {\eta}/{2} = \eta/2 >0,
\ee
where the second inequality is obtained by using Eq.~(\ref{eq:pu_q_guarantee}). When $\widehat{q}_l(I)-\widehat{p}_u(I)=0$, we have $\vert \widehat{p}_u(I) \vert >0$.  By Lemma \ref{lemma_two_cases_of_pu}, we see that $I$ is a maximal monomial of $p_u$.  
We have shown case 1).

For case 2), $\vert \widehat{q}_l(I) \vert < \eta$, by using again Eq.~(\ref{eq:pu_q_guarantee}), we can also show the upper bound
\be
\vert\widehat{p}_u(I) \vert = \vert \widehat{ p}_u(I)- \widehat q_l(I) + \widehat q_l(I) \vert 
\leq  \eta/2 +\eta < 2 \eta.
\ee
for the case that $I$ is a maximal monomial of $p_u-q$. For the case that $I$ with zero coefficient of $p_u-q$, we have $\vert \widehat{p}_u(I) \vert <\eta$. Since it is strictly less than $2\eta$ it cannot be a maximal monomial of $p_u$ because of the $\eta$-identifiable property. 

Take $I \subseteq [n]\setminus \{u\}$ for which
$\widehat{p}_u(I)-\widehat{q}_l(I)=0$. Then case 1) is true because $\widehat{p}_u(I)=\widehat{q}_l(I)\geq \eta >0$. Case 2) also true because
$\widehat{p}_u(I)=\widehat{q}_l(I)\leq \eta$ which implies $I$ is not a maximal monomial because of the $\eta$-identifiable property.  By Lemma \ref{lemma_two_cases_of_pu}, we see that $I$ is not a monomial of $p_u$.   
\end{proof}

\subsection{Main results of the modified structure learning algorithm for an r-wise MRF}

We construct a modified MRF structure learning algorithm based on the algorithm of Ref.~\cite{klivans2017learning}. As the original algorithm, we are able to recover the structure of the underlying graph of an $\eta$-identifiable and $r$-wise MRF, given $M$ samples of the MRF. 
The explanation for the algorithm is as follows. 
The algorithm has $r-1$ iterations compared to the single step of the algorithm of Ref.~\cite{klivans2017learning}.
The algorithm iterates over a parameter $l$ (from $r-1$ to $1$) which  corresponds to the size of the subsets considered in the current step of the loop.
We terminate this iteration early if  $\vert S \vert=n-1$, as there are at most $n-1$ neighbors of the vertex $u$.
Let 
\be
\bm p_{u,l} := \left(\widehat{p}_u(I): I\in[n]\setminus \{u\}, (\vert I \vert \leq l) \vee (I\in W_l) \right),
\ee
as in Lemma \ref{lemma_two_cases_of_pu}.
Note that $\bm p_{u,r-1} \triangleq \bm p_{u}$ with $\bm p_{u}$ of Eq.~(\ref{eqPuvec}),
which considers the coefficients of all potential monomials of size at most $r-1$.
We find all maximal monomials of $p_{u}$ with size exactly $r-1$. 
After doing so, we shrink the size as 
\be
\bm p_{u,r-1} \to \bm p_{u,r-2},
\ee 
by discarding the monomials of size $r-1$ with zero coefficient. 
Then, we find all the maximal monomials of size $r-2$, and discard the monomials of size $r-2$ with zero coefficient. In this way, we can find all the maximal monomials of the polynomial $p_u$.    

To find all maximal monomials of $p_{u,l}$ for each $l\in[r-1]$, we proceed as follows.
Let ${X}^{(m)}$ be the vector consisting of all monomials $Z_I^{(m)}=\prod_{k\in I}Z_k^{(m)}$.
We construct a vector $ \widetilde{X}^{(m)}=\{{X}^{(m)},-{X}^{(m)},0\}$ for each $m\in[M]$.
Taking $\widetilde{X}^{(m)}$ for all $m\in [M]$ as input,  apply the \textsc{Sparsitron} algorithm to obtain a vector $\bm{q}$ which is an estimate of $\bm p_{u,l}$. Then, we find all subsets $I$ with size $l$ which are not contained in any already found maximal monomials. 
We find the subset for which the coefficients satisfy $\vert \widehat{q}(I) \vert \geq \eta$. 
By using Lemma \ref{lemma_threshold_eta1}, we find all maximal monomial of $p_u$ with size $l$. 
Then the structure of the underlying graph of the MRF can be recovered by applying Algorithm \ref{alg_classical_MRF_sparsitron} for every vertex. The algorithm is shown in Figure \ref{alg_classical_MRF_sparsitron}.
The number of samples and the run time are given in the following theorem.

\begin{figure}[htbp]
\begin{algorithm}[H]
\caption{Iterative MRF Structure Learning Via Sparsitron (\textsc{MrfLearningIterative})}\label{alg_classical_MRF_sparsitron}
\begin{algorithmic}[1]
\Require{ $T+M$ samples on $Z\in\{1,-1\}^n$ from a $r$-wise MRF, 
$u\in[n]$,
$\eta >0, \lambda >0$, $r \in [n]$ }
\State Initialize  $J, S\gets\emptyset$, subset $I\subset [n]\setminus \{u\}$,  and $l \gets r-1$. 
\While{ $|S|<n-1$ and $l \neq 0$} \label{algorithm_classical_rbm_while}
\State $J_l' \gets J.$
\State\label{algorithm_classical_mrf_line_W} $W_l' \gets\{I \mid  \exists A \in J_l', I \subseteq A \wedge \vert I \vert >l  \}$.
\State $K_l \gets \sum_{k=1}^{l}  \tbinom{n-1}{k} +|W_l'|$.
\State\label{classical_mrf_step_input} For each $i\in [T+M],$ construct a vector $ \widetilde{X}^{(i)}=\left({X}^{(i)},-{X}^{(i)},0\right)$ with dimension $2K_l+1$,  and ${X}^{(i)}$ is the vector consisting of products $Z_I^{(i)}=\prod_{k\in I}Z_k^{(i)}$ for all subsets $I \in W_l'$ and all subsets $I$ where $\vert I \vert \leq l$, and $Y^{(i)}=\left( 1-Z^{(i)}_u\right)/2.$  
\State\label{algorithm_classical_mrf_line_sparsitron} 
$\bm {\widetilde q}_l \gets$ Apply the Sparsitron (Algorithm \ref{algosparsitron}) with input $$ \left(T,M, 2\lambda, (\widetilde X^{(i)},Y^{(i)})_{i=1}^{T+M}\right).$$
\State $\{\bm{q}_1, \bm{q}_2, q_3\}\gets \bm {\widetilde q_l}$, where  $\bm{q}_1$, $\bm{q}_2$ are $K_l$ dimensional subvectors.
\State $\bm{q}_l\gets \bm{q}_1- \bm{q}_2$. \label{algorithm_classical_mrf_line_polytransform}
 \For {each  $I \subset [n] \setminus \{u\}$ with  $|I|=l$ and $\left( \forall A \in J'_l, I \nsubseteq A\right)  $}
\label{algorithm_classical_rbm_for}
\State If $|\widehat q_l(I)| > \eta$, then
$J \gets J \cup \{I\}$, $S\gets S\cup I$.
\EndFor
\State $l \gets l-1$.
\EndWhile
\label{algorithm_classical_rbm_endwhile}
\Ensure{S}
\end{algorithmic}
\end{algorithm}
\end{figure}

 \begin{theorem} \label{th:identifiable}
Let $\cal{D}$ be a  $r$-wise MRF on $\{1,-1\}^n$  with underlying graph $G$ and factorization polynomial $p(Z) = \sum_{I \in C_r(G)} \widehat p(I)Z_I$ with $\max_i \Vert\partial_i p\Vert_1 \leq \lambda$. With $\eta>0$, assume that $\cal{D}$ is $\eta$-identifiable. Then given $\lambda$, $\rho \in (0,1)$, and 
$$M = {e^{O(r)} e^{O( \lambda r)}} \log (nr/\rho \eta)/{\eta^4}$$ independent samples from $\cal{D}$, by using Algorithm \ref{alg_classical_MRF_sparsitron} for every vertex, the structure of the underlying graph $G$ can be recovered in time  $\Ord{Mn^r}$  with probability at least $1-\rho$.
\end{theorem}
\begin{proof}  
In the first step of the loop in Algorithm \ref{alg_classical_MRF_sparsitron}, note that trivially $J'_{r-1} = J_{r-1}(p_u)=\emptyset$ and $W'_{r-1} = W_{r-1}(p_u)=\emptyset$.
For $r-1$, after Line \ref{algorithm_classical_mrf_line_polytransform},  we obtain a polynomial  
$$q_{r-1}=\sum_{|I|\leq l}\widehat{q}_{r-1}(I)Z_I,$$ 
which is an estimation of the polynomial $p_{u,r-1}$. Using  Lemma \ref{lemma_two_cases_of_pu} and Lemma \ref{lemma_threshold_eta1} we can show that $J'_{r-2}= J_{r-2}(p_u)$, i.e., we have found all maximal monomials of $p_u$ with size exactly $r-1$, with the success probability given from the Sparsitron. 

Now assume we are at step $l
\in [r-1]$ and we prove for that we obtain the valid sets for $l-1$. Assume the induction hypothesis holds that $J'_{l} = J_{l}(p_u)$ holds. Hence, also $W'_{l} = W_{l}(p_u)$ holds. 
After Line \ref{algorithm_classical_mrf_line_polytransform},  we obtain a polynomial  $$q_l =\sum_{|I|\leq l}\widehat{q}_l(I)Z_I+\sum_{I\in W_l(p_u)}\widehat{q}_l(I)Z_I,$$ which is an estimation of polynomial $p_{u,l}$. As shown in Lemma \ref{lemma_two_cases_of_pu}, for a subset  $I \subset [n]\setminus \{u\} $ with size $l$,  if it is not a subset of any  element of $J_l$, it is either a  maximal monomial of $p_{u,l}$ or with zero coefficient in $p_{u,l}$. It is also a maximal monomial of $p_{u,l}-q_l$ if $\widehat{p}_{u,l}(I)-\widehat{q}_l(I) \neq 0$. According to  Lemma \ref{lemma_threshold_eta1}, we can infer whether a subset $I$ is a maximal monomial of $p_{u,l}$ from the value of $\vert  \widehat q_l(I) \vert $. 
The Sparsitron step succeeds with probability $1-\rho/(n(r-1))$. In case it succeeds, we have that $J'_{l-1}= J_{l-1}(p_u)$, i.e., we have found all maximal monomials of $p_{u,l}$ of size exactly $l$.
This proves the induction step. 
Hence, at the end, we obtain all maximal monomials of $p_{u}$ and all the neighbors of vertex $u$.

By Theorem \ref{algosparsitron}, the number of samples required for each call to the Sparsitron Algorithm in Line \ref{algorithm_classical_mrf_line_sparsitron} of  Algorithm \ref{alg_classical_MRF_sparsitron} is given by
\begin{eqnarray}
\Ord{\lambda^2(\log(rn^r/\rho \epsilon))/\epsilon^2} = e^{\Ord{r}}  e^{\Ord{\lambda r}}{\log(nr/\rho \eta)/ \eta ^4}, 
\end{eqnarray}
 since $\Vert p_u \Vert_1 \leq 2\lambda$, $\delta= (e^{-2\lambda}/2)$ and $\epsilon = \Ord{e^{-2\Vert p_u \Vert_1 -6} \delta^{r} \eta^2}$ by Lemma \ref{lemma_threshold_eta0}.
 
Now we analyze the run time.  In Algorithm \ref{alg_classical_MRF_sparsitron}, for each $l$ from $r-1$ to $1$, there are at most $K_l \triangleq\sum_{k=1}^{l}  \tbinom{n-1}{k} +|W_l| =\Ord{n^{l}+|W_l| }$ potential monomials of polynomial $q_l$ containing vertex $u$. Notice that $\Ord{n^{l}+|W_l|}$ is bounded by  $\Ord{n^{r-1}}$.  For each loop, Line \ref{algorithm_classical_mrf_line_sparsitron}
is the most time-consuming step which costs at most $\Ord{n^{r-1} M}$ run time as it calls the Sparsitron Algorithm. The run time of the \emph{for} loop is bounded by  $\Ord{n^{r-1}}$ as the number of potential monomials is bounded by $\Ord{n^{r-1}}$. Then the total run time of Algorithm \ref{alg_classical_MRF_sparsitron} is $\Ord{rn^{r-1} M}$ since it runs at most $r-1$ times from Line \ref{algorithm_classical_rbm_while} to \ref{algorithm_classical_rbm_endwhile}. 
Hence, it results in time $\Ord{rn^r M}$ over all $n$ vertices. As the factor $r$ is contained in the factor $e^{\Ord{r}}$ of $M$ ($re^{\Ord{r}}$ is  bounded by $e^re^{\Ord{r}}$ and $e^re^{\Ord{r}}=e^{r+\Ord{r}} = e^{\Ord{r}}$). The run time is then $\Ord{n^r M}$.

For $r-1$ runs of the Sparsitron algorithm,  the success probability is bounded by $1-\frac{\rho}{n}$ with Boole's inequality. Run Algorithm \ref{alg_classical_MRF_sparsitron} for $n$ vertices, the total success probability is then bounded by $1-\rho. $
\end{proof}

\section{Quantum set membership queries for MRF structure learning}
\label{secqsetmrf}

Based on the classical Algorithm \ref{alg_classical_MRF_sparsitron}, we would like to construct a quantum algorithm to learn the structure of MRFs with quantum advantage. The quantum algorithm requires certain sets of strings and corresponding data structures which are introduced in this section. These string sets are analogues to the sets of the previous sections. The data structures rely on the availability of quantum RAM and allow quantum set membership queries as discussed in Theorem \ref{thmQSet_informal} and Appendix \ref{appendix_quantum_indicator_function}. 
We first discuss four sets of strings that correspond to four types of subsets of the nodes of the MRF (again the class of MRFs is given by Definition \ref{defMRFclass}). For some of the sets, we give the run time of setting up data structures and the quantum set membership query. This query allows to determine if a given string is an element of a given set. 

Fix a node $u$ and consider the derivative polynomial $p_u$.
For the $r$-wise MRF, each monomial in $p_u$ involves at most $r-1$ vertices, since  $u$ is excluded.
The following discussion pertains to any multi-linear polynomial, however, we choose 
$p_u$ and its approximation $q$ with $r-1$ variables excluding $u$ as the basis of the discussion to keep the connection to the MRF.

As before, we associate every monomial by the indices of the variables contained in the monomial. It is beneficial for the quantum algorithm to work with fixed-length strings that describe the variables contained in a monomial. There is a one-to-many mapping of subsets to strings, since many strings can describe the same subset. 
More formally, define strings $\mathcal{S}$ of a certain length (here $r-1$)  where each element in a string is chosen from $\{0\} \cup [n]$. 
Here, $0$ is used as a padding element.
Each string can be mapped to a subset  $I\subset[n]$ with size $\vert I \vert \leq r-1$,
where we ignore elements $0$ in the string. 
\begin{defn}[String and the corresponding  subset] \label{defString}
A string of length $r-1$ is defined as $\mathcal S:=(j_1,\cdots, j_{r-1}) \in (\{0\} \cup [n])^{r-1}$.
The corresponding subset is defined as the set consisting of the non-zero elements of a string $\mathcal{S}$, which is denoted as $I(\mathcal S)$.
\end{defn}
For example,  if $\mathcal{S}=(1,3,2,3,4,0,5)$, then $I(\mathcal{S})=\{1,2,3,4,5\}$.
A string $\mathcal{S}$ can be represented by a quantum state with $(r-1)\lceil \log (n+1) \rceil \in \Ord{r\log n}$ qubits. The ``string quantum state" is defined as the following.
\begin{defn}[String quantum state]\label{defStringQuantumState}
Let $j_1,\cdots, j_{r-1} \in [n]\cup \{0\}$ and the corresponding string be $\mathcal{S}_j=(j_1,\cdots, j_{r-1})$. Define the one-to-one shorthand notation for the $\Ord{r \log n}$ qubit state
\be
\ket{j_1}\ket{j_2}\cdots \ket{j_{r-1}} =: \ket {\mathcal{S}_j}.
\ee
\end{defn}
Next, we consider strings that exclude a certain vertex $u$ (due to the connection to the polynomial $p_u$) and have other properties. 
For each $l\in[r-1]$, we define four types of sets $\mathcal H_l$, $\mathcal J_l$, $\mathcal W_l$ and $\mathcal F_l$ consisting of strings of size $r-1$, and study the quantum set membership query for sets $\mathcal H_l$, $\mathcal W_l$, and $\mathcal F_l$.

Many different strings may map to the same subset. 
For strings in $\mathcal H_l$, we associate subsets of nodes with sizes not larger than $l$. In addition, the strings in $\mathcal H_l$ are defined such that the non-zero elements are all different and sorted in ascending order, with the zeros at the end. 
\begin{defn}[Ordered strings]\label{SubsetH}
Define $\mathcal H_0 =\emptyset$.
For fixed $u\in[n]$ and for all $l\in [r-1]$,
define the set $\mathcal H_{l}$ as set of strings $\mathcal S$ with the following conditions:
\begin{itemize}
\item[•] For all $\mathcal S \in \mathcal H_l$, $u \not \in I(\mathcal S)$.
\item[•] For all $\mathcal S \in \mathcal H_l$, $0<\vert I(\mathcal S) \vert \leq l$.
\item[•] For all $\mathcal S\in \mathcal H_l$, there exists a $k\in [l]$ such that $0< j_1 < \cdots < j_{k}$, and $j_{k+1}=\cdots =j_{r-1}=0$ (the latter condition applies only if $k<r-1$).
\end{itemize}
\end{defn}
As an example, for $n=4$, $r =3$, $l=2$, $u=1$, we have the string set $\mathcal H_2 = \{ (2, 0), (3, 0), (4, 0), (2, 3), (2, 4), (3, 4) \}.$ 

\begin{lemma}\label{lemma_indicator_subsets_Hl}
For the sets  $\mathcal H_l$, there is a unitary $U_{\mathcal H_l}$ which performs the quantum set membership query (quantum indicator function) 
\begin{eqnarray}
 \ket{\mathcal{S}_j }\ket{0}\to
\begin{cases}
 \ket{\mathcal{S}_j }\ket{1} &\text{for} ~  \mathcal{S}_j \in \mathcal H_l\\
 \ket{\mathcal{S}_j } \ket{0}, & \text{for} ~ \mathcal{S}_j \notin \mathcal H_l,
\end{cases}
\end{eqnarray}
in time  $\Ord{r\log n }$.
\end{lemma}
\begin{proof}
Given Definition \ref{SubsetH}, the following steps require at most $\Ord{r}$ comparisons, and each comparison involves $\Ord{\log n}$ qubits.
Consider an ancillary output register of size $l+2$.
First, output $0$ to the first position of the ancillary output register if any one of $j_1,\cdots, j_{r-1}$ is equal to $u$, and $1$ otherwise.
Then, output $0$ to the second position of the ancillary output register if any one of $j_{l+1}, \cdots, j_{r-1}$ is greater than $0$, and $1$ otherwise.
For each $i \in [l]$, output $0$ to the $i+2$-th position of the ancillary register if $j_i \ge j_{i+1}$, or $j_i = 0$ and $j_{i+1} > 0$, and output $1$ otherwise.
For each of these steps, we have $\Ord{1}$ comparisons.
We have in total $l+2$ output ancilla registers. If there is any $0$ in those registers, output $0$ to the result register, otherwise output $1$ to the result register. 
Finally, uncompute the ancillary register and keep the result register.
Hence, the indicator function costs $\Ord{r \log n}$ to compute.
\end{proof}

The next set $\mathcal J_l$ corresponds to the set $J_l$ in Definition \ref{Subset_classical_JWF} but for the strings. It is a subset of  $\mathcal H_{r-1}\setminus \mathcal H_l$ which contains strings with at least $l+1$ non-zero elements. It indicates that the size of subsets corresponding to strings in $\mathcal J_l$ is larger than $l$. 
Additionally, the set $\mathcal J_l$ is again defined in relation to any multi-linear polynomial $q$.
It holds that $J_l = I(\mathcal J_l)$.

\begin{defn}\label{SubsetJ}
For all $u\in [n]$, define $\mathcal J_{r-1} = \emptyset$.
For all $l\in [r-2]$, $u\in [n]$,  and let there be given string sets 
$\mathcal H_{r-1}$ and $\mathcal H_{l}$ as in Definition \ref{SubsetH}, and any multi-linear polynomial $q$ on variables indexed by $[n]\setminus \{u\}$.
Define the set $\mathcal J_l$ as the set of the strings $\mathcal S$ with the following conditions:
\begin{itemize}
\item[•] For all $\mathcal S \in \mathcal J_l $,  we have that $\mathcal S \in \mathcal H_{r-1}\setminus \mathcal H_l$, which implies  $l<\vert I(\mathcal S)\vert \leq r-1$. 
\item[•] For all $\mathcal S \in \mathcal J_l $, we have $I(\mathcal S)$ is a maximal monomial of $q$.
\end{itemize}
\end{defn}
For example, given $n=4$, $r=3$, $u=1$, and $l=1$, we have $\mathcal H_2\setminus \mathcal H_1= \{(2, 3), (2, 4), (3, 4) \}$. 
Let a multi-linear polynomial be $q_{\rm ex} := 2 z_2 z_4 + 0.5 z_2  + 0.3 z_3$. We obtain $\mathcal J_1 = \{ (2,4 )\}$.
Now we define a subset $\mathcal W_l$, 
for which $W_l = I(\mathcal W_l)$.
\begin{defn}\label{SubsetW}
Fix $l\in [r-1]$, given a string set $\mathcal J_l$ as in Definition \ref{SubsetJ}. Define the set $\mathcal W_l$ as the set of the strings $\mathcal S$ with the following conditions:
\begin{itemize}
\item[•] For all $\mathcal S \in \mathcal W_l $, we have that $\mathcal S \in \mathcal H_{r-1}\setminus \mathcal H_l$, which implies $l<\vert I(\mathcal S)\vert \leq r-1$. 
\item[•] For all $\mathcal S \in \mathcal W_l $,  $\exists \mathcal A \in \mathcal J_l$ such that $ I(\mathcal S) \subseteq I(\mathcal A)$. 
\end{itemize}
\end{defn}
In the example after Definition \ref{SubsetJ}, $\mathcal W_1 = \mathcal J_1$. Let $\mathcal J_l(q), \mathcal W_l(q)$ denote the string set $\mathcal J_l, \mathcal W_l$ referring to polynomial $q$.
We have the following size bound and quantum access for these sets.
\begin{lemma}\label{lemma_indicator_subsets_Wl}
For the sets $\mathcal J_l$ and $\mathcal W_l$ from Definitions \ref{SubsetJ} and \ref{SubsetW}, the following holds.
\begin{enumerate}[1)]
\item If the polynomial $q$ in Definition \ref{SubsetJ} is over $d < n$ variables and each monomial of $q$ contains at most $r-1$ variables, then $\vert \mathcal J_l\vert \leq \vert \mathcal W_l\vert \leq \mu$, where $\mu := \min (2^d-1,d^{r-1})$. 
\item A quantum data structure for $\mathcal W_l$ can be constructed in time $\tOrd{  \mu r \log n}$, 
such that there is a unitary $U_{\mathcal W_l}$  which performs the following quantum indicator function (or set membership query)
\begin{eqnarray}
 \ket{\mathcal{S}_j }\ket{0}\to
\begin{cases}
 \ket{\mathcal{S}_j }\ket{1} &\text{for} ~  \mathcal{S}_j \in \mathcal W_l\\
 \ket{\mathcal{S}_j } \ket{0}, & \text{for} ~ \mathcal{S}_j \notin \mathcal W_l
\end{cases}
\end{eqnarray}
in time $\Ord{r\log n\log \mu+\log^3 \mu}$. 
\end{enumerate}

\end{lemma}
\begin{proof}
For 1), we always have that $\vert \mathcal J_l\vert \leq \vert \mathcal W_l\vert$ since  $\mathcal J_l$ is a subset of $\mathcal W_l$.
By assumption the polynomial $q$ is a polynomial over at most $d < n$ variables.
We also know that its degree is at most $r-1$. 
(Of course $r-1\leq d$.) 
Given these assumptions, the number of monomials in $q$ is at most $\sum_{i=1}^{r-1}\tbinom{d}{i}$.
This number is bounded by $\mu := \min (2^d-1,d^{r-1})$, hence the size of $ \mathcal W_l$ is  bounded by $\mu$.

For 2), by using  Lemma \ref{QRAM_construction}, it requires at most $\tOrd{\vert\mathcal W_l\vert r\log n}$ time to construct a sorted QRAM for $\mathcal W_l$. According to  Theorem \ref{Quantum_indicator_function}, it requires time $\Ord{\log \vert \mathcal W_l\vert \log(n^{r-1}) +\log^3 \vert \mathcal W_l \vert } \subseteq \Ord{r\log n\log \mu+\log^3 \mu}$
to implement unitary $U_{\mathcal W_l}.$
\end{proof}

We now show the simple operation of flagging ordered strings as defined by the union of the sets $\mathcal H_l$ and $\mathcal W_l$ in the following lemma. This operation will be needed below in Lemma \ref{lemma_sparsitron_input}.

\begin{lemma}\label{lemma_indicator_hl_wl}
Fix a node $u\in [n]$, given the sets $\mathcal H_l$ and $\mathcal W_l$ as defined in Definitions \ref{SubsetH} and \ref{SubsetW}.  After constructing the QRAM for $\mathcal W_l$ in time $\tOrd{\mu\log n}$, where $\mu= \min (2^d-1,d^{r-1})$,  for string $\mathcal{S}_j=(j_1,\cdots j_{r-1}) \in ([n] \cup \{0\})^{r-1}$,  there is a unitary $U_{\mathcal H_l \cup \mathcal W_l}$ which performs
\begin{eqnarray}
\ket{0}\ket{\mathcal{S}_j}
\to 
\begin{cases}
\ket{0} \ket{\mathcal{S}_j},  & ~\text{for} ~\mathcal{S}_j\notin (\mathcal H_l\cup \mathcal W_l) \\
\ket{1} \ket{\mathcal{S}_j},   & ~\text{for} ~ \mathcal{S}_j \in  (\mathcal H_l \cup \mathcal W_l)
\end{cases} \label{eq_indicator_Hl_Wl}
\end{eqnarray}
with $\Ord{r \log n \log\mu +\log^3\mu}$ quantum gates.
\end{lemma}
\begin{proof}
 Given set $\mathcal H_{l}$ and set $\mathcal W_l$ as  Definitions \ref{SubsetH} and \ref{SubsetW} and a sorted QRAM for set $\mathcal W_l$,  we can obtain 
\begin{eqnarray}
 \ket{\mathcal{S}_j} \ket{0}_a \ket{0}_b \to 
 \begin{cases}
\ket{\mathcal{S}_j}\ket{1}_a \ket{0}_b, ~~&\text{for}~~  \mathcal{S}_j \in \mathcal H_l\\
\ket{\mathcal{S}_j}\ket{0}_a \ket{1}_b, ~~&\text{for}~~  \mathcal{S}_j \in \mathcal W_l\\
 \ket{\mathcal{S}_j} \ket{0}_a\ket{0}_b,~~&\text{for}~~  \mathcal{S}_j\notin (\mathcal H_l \cup \mathcal W_l) 
\end{cases}
\end{eqnarray}
by performing  the Unitary  $U_{\mathcal H_{l}}$ on states $ \ket{\mathcal{S}_j} \ket{0}_a $  and $U_{\mathcal W_l}$ on states $  \ket{\mathcal{S}_j} \ket{0}_b$. There is no result $\ket{\mathcal{S}_j} \ket{1} \ket{1}$ because the intersect of $\mathcal H_l$ and $\mathcal W_l$ is empty. 
Hence we can identify a string in $\mathcal H_l \cup \mathcal W_l$ by $\ket{0}_a\ket{1}_b$ and $\ket{1}_a\ket{0}_b$. We output the result into the first register in Eq.~$( \ref{eq_indicator_Hl_Wl})$.

For the run time, as shown in Lemma \ref{lemma_indicator_subsets_Wl}, constructing the sorted QRAM for set $\mathcal W_l$ requires time $\tOrd{\mu \log n}$ and performing  the unitary  $U_{\mathcal W_l}$  costs $\Ord{r\log n \log\mu +\log^3\mu }$ run time. By Lemma \ref{lemma_indicator_subsets_Hl},  performing  the unitary  $U_{\mathcal H_{l}}$  costs  $\Ord{r\log n }$ run time. Thus the run time for applying  unitary $U_{\mathcal H_l\cup \mathcal W_l}$  is $\Ord{r \log n \log\mu+\log^3\mu }$.
\end{proof}

For the next set $ I(\mathcal F_l)=F_l$, analogous to the other sets before. 
\begin{defn}\label{SubsetF} 
Fix  $l\in [r-2]$ and a multi-linear polynomial $q$, which fixes $\mathcal J_l$ from Definition \ref{SubsetJ}. Given $\mathcal H_l$ in Definition \ref{SubsetH} (where $\mathcal H_0=\emptyset$), define the set $\mathcal F_l$ as the set of the strings $\mathcal S$ with the following conditions:
\begin{itemize}
\item For all $\mathcal S \in \mathcal F_l $, we have  $\mathcal S \in \mathcal H_l \setminus \mathcal H_{l-1}$, which implies $\vert I( \mathcal S)\vert = l$.
\item For all $\mathcal S \in \mathcal F_l $, we have   $ \forall \mathcal A \in \mathcal J_l, I(\mathcal S)  \nsubseteq I(\mathcal A)$.
\end{itemize}
Note that we also define the complement of $\mathcal F_l$ inside the boundary $\mathcal H_l \setminus \mathcal H_{l-1}$ as 
$\bar{\mathcal F}_l := (\mathcal H_l \setminus \mathcal H_{l-1}) \setminus \mathcal F_l$, for which each string is a substring of $\mathcal J_l$ (the negation of bullet 2).
\end{defn}
We have the following size bound and quantum access  for these sets.
\begin{lemma}\label{lemma_indicator_subsets_Fl}
If the polynomial $q$  contains at most $d$ variables, for the subsets $\mathcal F_l$ and $\bar{\mathcal F}_l$ in Definition \ref{SubsetF} the following holds. 
\begin{enumerate}[1)]
\item $\vert \bar{\mathcal F}_l \vert \leq \kappa$ with $\kappa =\min (2^d/\sqrt{d}, d^{r-2})$,
\item A sorted QRAM for $ \bar{\mathcal F}_l $ can be constructed in time $\tOrd{ \kappa \log n } $, such that there is a unitary $U_{{{\mathcal F}_l} }$ performing the following quantum indicator function (or set membership query)
\begin{eqnarray}
 \ket{\mathcal{S}_j }\ket{0}\to
\begin{cases}
 \ket{\mathcal{S}_j }\ket{1} &\text{for} ~  \mathcal{S}_j \in \mathcal F_l\\
 \ket{\mathcal{S}_j } \ket{0}, & \text{for} ~ \mathcal{S}_j \notin \mathcal F_l
\end{cases}
\label{Equ_Fl}
\end{eqnarray}
in time $\Ord{r \log n \log \kappa +\log^3\kappa } $ for all $l\in[r-2]$.
\end{enumerate}
\end{lemma}
\begin{proof}
For 1), as there are at most  $\tbinom{d}{l}$ monomials  of size $l$ in polynomial $q$,  the size of set $\bar{\mathcal F}_l$ is at most $\tbinom{d}{l} \leq \min( \tbinom{d}{\lfloor d/2\rfloor },  d^{r-2} )$.  It is bounded by $\kappa =\min (2^d/\sqrt{ d}, d^{r-2})$ as  $\tbinom{d}{\lfloor d/2\rfloor } < 2^d/\sqrt{ d}$.
For 2), there is a unitary $U_{\mathcal H_l \setminus \mathcal H_{l-1} }$ which performs, 
\begin{eqnarray}
 \ket{\mathcal{S}_j }\ket{0}\to
\begin{cases}
 \ket{\mathcal{S}_j }\ket{1} &\text{for} ~  \mathcal{S}_j \in \left(\mathcal H_l \setminus \mathcal H_{l-1} \right)\\
 \ket{\mathcal{S}_j } \ket{0}, & \text{for} ~ \mathcal{S}_j \notin \left( \mathcal H_l \setminus \mathcal H_{l-1} \right),
\end{cases}
\end{eqnarray}
in time $\Ord{r\log n}$. 
To do this, return $0$ if one of $j_1,\cdots, j_{r-1}$ is equal to $u$. 
Output $1$ in a distinct register if $ j_i < j_{i+1}$ for all $i<l$ and if $j_{l+1},\cdots, j_{r-1}$ are all equal to zero.
Next, construct a sorted QRAM for the strings in  set  $\bar{\mathcal F}_l $.
For a string $\mathcal{S} \in \bar{\mathcal F}_l $, the corresponding set $ I(\mathcal{S})$ is a subset of certain element in $I(\mathcal J_l)$.  According to Theorem \ref{Quantum_indicator_function},  there is a unitary operator $U_{{ \bar{\mathcal F}_l}}$  which performs 
\begin{eqnarray}
 \ket{\mathcal{S}_j }\ket{0}\to
\begin{cases}
 \ket{\mathcal{S}_j }\ket{1} &\text{for} ~  \mathcal{S}_j \in \bar{\mathcal F}_l\\
 \ket{\mathcal{S}_j } \ket{0}, & \text{for} ~ \mathcal{S}_j \notin \bar{\mathcal F}_l.
\end{cases}
\end{eqnarray}
Now we implement the desired quantum operation. Start with state $ \ket{\mathcal{S}_j }\ket{0}_a\ket{0}_b$,  apply unitary  $U_{{\mathcal H_l \setminus \mathcal H_{l-1}}}$ on $ \ket{\mathcal{S}_j }\ket{0}_a$ then  $U_{\bar{\mathcal F}_l}$ on $ \ket{\mathcal{S}_j }\ket{0}_b$.
This results in
\begin{eqnarray}
 \ket{\mathcal{S}_j }\ket{0}_a\ket{0}_b \to
\begin{cases}
 \ket{\mathcal{S}_j }\ket{1}_a\ket{0}_b \  &\text{for} ~  \mathcal{S}_j \in \mathcal F_l\\
 \ket{\mathcal{S}_j }\ket{1}_a\ket{1}_b  & \text{for} ~ \mathcal{S}_j \in \bar{\mathcal F}_l\\
  \ket{\mathcal{S}_j }\ket{0}_a\ket{0}_b  & \text{for} ~ \mathcal{S}_j \notin (\mathcal H_l \setminus \mathcal H_{l-1}). 
\end{cases}
\end{eqnarray}
Hence, we uniquely identify $\mathcal F_l$ by the string $\ket{1}_a\ket{0}_b$, and we can output the corresponding bit into the output register.
Undoing the unitary $U_{\bar{\mathcal F}_l}$  and the unitary $U_{{(\mathcal H_l \setminus \mathcal H_{l-1})}}$ disentangles qubits $a,b$ and yields the result in Eq.~(\ref{Equ_Fl}). 
 
By Lemma \ref{QRAM_construction}, to construct a QRAM for set  $\bar{\mathcal F}_l$, it costs at most $\tOrd{r\kappa \log n}$   run-time. As $r=\tOrd{\log \kappa}$, we have $\tOrd{ r\kappa \log n }=\tOrd{\kappa \log n}$.  According to Theorem \ref{Quantum_indicator_function}, it requires $\Ord{r \log n \log \kappa +\log^3\kappa}$ run-time to perform $U_{\bar{\mathcal F}_l}$. Combing with the time $\Ord{r\log n}$ for implementing $U_{\mathcal H_l \setminus \mathcal H_{l-1}}$, it requires $\Ord{r\log n \log \kappa +\log^3\kappa }$ to perform unitary $U_{{\mathcal F_l}}$. 
\end{proof}

\section{Quantum MRF structure learning algorithm}
\label{section_quantum_mrf_algorithm}

Based on the classical Algorithm \ref{alg_classical_MRF_sparsitron}, we construct a quantum algorithm to learn the structure of $r$-wise,  $\eta$-identifiable MRFs. Recall that  structure learning obtains the neighbors of each node of the underlying graph. 
We have the additional assumption that  the degree of the underlying graph is bounded by $d$, whereas there is no degree restriction in the classical algorithm,
and that we are given quantum access to a number of samples of the MRF.
The quantum algorithm  provides a polynomial speedup over the classical Algorithm in terms of the dimension of the samples.

\subsection{Quantum data input and monomials}

First of all, we introduce the quantum data input.  Assume that we have quantum access to the samples defined as follows. Here, by usual conventions, we do not assume anything about the run time of this oracle. In practice, such an oracle may be implemented via quantum RAM, which incurs further factors to the run time.
\begin{defn}[Quantum access to distribution samples]\label{MRF_input}
Given quantum access to $M$ samples from a distribution $\mathcal D$ on $\{-1,1\}^n$, denoted by $Z^{(1)},...,Z^{(M)} \in \{-1,1\}^n,$ for $j\in[n]$ and $m\in [M]$ we have
\begin{eqnarray}
\ket{j}\ket{0}\rightarrow 
~\ket{j}\ket{Z_j^{(m)}},
\end{eqnarray}
where $Z_j^{(m)} $ is the value of node $j$ in sample $Z^{(m)}$.
The $Z_j^{(m)}$ is simply stored in the state of a single qubit representing the value $\{-1,1\}$.
\end{defn}

We employ all the sets and set membership quantum queries discussed in the previous section. 
Given a string $\mathcal{S}=(j_1,\cdots,j_{r-1})$, we show that  the corresponding monomial excluding the $j_k$ that are zero  can be computed when given this quantum access.

\begin{lemma}[]\label{Lemma_Sparsitron_input1}
 Given quantum access to $M$ samples as in Definition \ref{MRF_input}. For each $m\in [M]$, there is the unitary $U_{Z}^{(m)}$ that for all $\mathcal{S}_j=(j_1,\cdots, j_{r-1})$, for $j_k \in[n]\cup \{0\}, \forall k\in[r-1]$, performs 
\begin{eqnarray} \label{eq_subset_mrf_sparsitron}
U_{Z}^{(m)} \ket{\mathcal{S}_j} \ket{\bar{0}}& =&
\ket{\mathcal{S}_j}  \ket{Z^{(m)}_{\mathcal{S}_j}},  
\end{eqnarray}
where $Z^{(m)}_{\mathcal{S}_j}:=\prod_{\substack{k\in \{k \vert k\in [r-1], j_k \neq 0 \} }} Z^{(m)}_{j_k }$ is the corresponding monomial.
The unitaries require 
 $\Ord{r}$ queries to the quantum access and $\Ord{r \log n}$ quantum gates.
The register $\ket{\bar{0}}$ consists of two qubits such that we are able to store the values $\{-1,0,1\}$.
\end{lemma}

\begin{proof}
For each sample $m\in [M]$ the following can be performed. Let $k \in[r-1]$ and  $j_k\in [n]$,  we can prepare, conditioned on $j_k > 0$,
\begin{eqnarray}
\ket{j_k}\ket{\bar{0}} \rightarrow \ket{j_k}\ket{Z_{j_k}^{(m)}}\label{eq_index_nonzero}
\end{eqnarray}
via quantum access from Definition \ref{MRF_input}. Conditioned on $j_k=0$, we prepare 
\begin{eqnarray}
\ket{j_k}\ket{\bar{0}} \rightarrow \ket{j_k}\ket{1},\label{eq_index_zero}
\end{eqnarray}
with $\Ord{1}$ queries and $\Ord{\log n}$ quantum gates.  For $r-1$ quantum registers, this process requires  $\Ord{r}$ queries and $\Ord{r\log n}$  quantum gates.  
We obtain a binary representation of $Z^{(m)}_{\mathcal{S}_j} \equiv \prod_{\substack{k\in \{k \vert k\in [r-1], j_k \neq 0 \} }} Z_{j_k}^{(m)}$ by multiplying the results of the query of Eq.~(\ref{eq_index_nonzero}) and Eq.~(\ref{eq_index_zero}).  Then undoing the oracles and the operation in Eq.~(\ref{eq_index_zero}), we have 
$\ket{\mathcal{S}_j}\ket{Z^{(m)}_{\mathcal{S}_j}}$.
\end{proof}
Notice that if $\mathcal{S}_j$ above is a string with non-repeating non-zero elements, we see that $Z^{(m)}_{\mathcal{S}_j} = Z^{(m)}_{I(\mathcal{S}_j)}$.  Therefore, the lemma can be used to calculate the multi-linear monomials of a given MRF. 

\subsection{Quantum Sparsitron  with MRF quantum data input}

The Sparsitron is the main subroutine of the classical algorithm. We use the quantum Sparsitron algorithm discussed in  Ref.~\cite{rebentrost2021quantum} as a subroutine of our quantum MRF structure learning algorithm.   
The quantum Sparsitron algorithm is shown in Algorithm \ref{algSparsitronQuantum} in Appendix  \ref{apendix_quantum_sparsitron}. As shown in Theorem \ref{theoremQuantumSparsitron}, for $N$-dimensional samples, in terms of only the dimension and no other parameters, the run time of the quantum Sparsitron algorithm is $\tOrd{\sqrt{N}}$ with   $\Ord{\log N}$ required samples.

We now construct the input for the quantum Sparsitron. 
Fix $l\in[r-1]$ and a set $\mathcal H_l$ and a set $\mathcal W_l$, recall Definitions \ref{SubsetH} and \ref{SubsetW}. For each $m \in [M]$, let ${X}^{(m)}$ be the vector consisting of all $Z_{\mathcal{S}_j}^{(m)}$, where $\mathcal{S}_j$ is a string in the union set of $\mathcal H_l$ and $\mathcal W_l$.  
As for the classical algorithm, to adapt to the Sparsitron algorithm which operates on positive weight vectors, we extend ${X}^{(m)}$ by adding a negative copy of the vector ${X}^{(m)}$ and $0$ such that each element in the new weight vector is non-negative and the $1$-norm of the weight vector is equal to exactly $2\lambda$ (for $\lambda > 0$). For the same reason,
in the quantum algorithm, we construct a quantum state which represents the vector containing the original input ${X}^{(m)}$, a negative copy of the original input, and also a number of zeros. We use these states as the input to the quantum Sparsitron algorithm.

In the following lemma, we construct a unitary, then apply it to a state in Lemma \ref{lemma_quantum_mrf_input}, which shows that we can prepare the input for the quantum Sparsitron algorithm.

\begin{lemma}\label{lemma_sparsitron_input}
Fix $l\in[r-1]$ and a set $\mathcal H_l$ and a set $\mathcal W_l$ as defined in Definition \ref{SubsetH} and \ref{SubsetW}.
Let  $m\in[M]$. Assume we are given the  controlled version of $U_{Z}^{(m)}$ from Lemma \ref{Lemma_Sparsitron_input1}, QRAM for $\mathcal W_l$, and the unitary $U_{\mathcal H_l\cup \mathcal W_l}$ in Lemma \ref{lemma_indicator_hl_wl}.
Then, there is a unitary $U_{\rm in}^{(m)}$ which 
 for all $\mathcal{S}_j=(j_1,\cdots j_{r-1}) \in ([n] \cup \{0\})^{r-1}$ performs
\begin{eqnarray}
\ket{0}\ket{+}\ket{\mathcal{S}_j}  \ket{\bar{0}}
\to 
\begin{cases}
\ket{0}\ket{+}\ket{\mathcal{S}_j}  \ket{\bar{0}},  & ~\text{for} ~\mathcal{S}_j\notin (\mathcal H_l\cup \mathcal W_l) \\
\frac{1}{\sqrt{2}} \left(\ket{1} \ket{0} \ket{\mathcal{S}_j}   \ket{Z_{\mathcal{S}_j}^{(m)}}+
\ket{1} \ket{1} \ket{\mathcal{S}_j}   \ket{-Z_{\mathcal{S}_j}^{(m)}} \right),   & ~\text{for} ~ \mathcal{S}_j \in  (\mathcal H_l \cup \mathcal W_l)
\end{cases} \nonumber\\
\label{eq_quantum_mrf_input}
\end{eqnarray}
with $\Ord{r\log n\log \mu+\log^3\mu }$ quantum gates.
\end{lemma}
\begin{proof}
We discuss for all states $\ket{0}\ket{+}\ket{\mathcal{S}_j}  \ket{\bar{0}}$. From Lemma \ref{lemma_indicator_hl_wl}, we first apply a unitary $U_{\mathcal H_l \cup \mathcal W_l}$ on the third and first registers.
Followed by applying a controlled $U_{Z}^{(m)}$ defined in Lemma \ref{Lemma_Sparsitron_input1} which is controlled by the first register and outputs on the target (last register). Then apply a controlled gate C-sign which  maps $\ket{0}\ket{b}\to\ket{0}\ket{b}$ and $\ket{1}\ket{b}\to\ket{1}\ket{-b}$. The gate is applied on the second register ($\ket{+}$) and the last register (controlled by the second register). The procedure is shown in the following
\begin{eqnarray}
&\ket{0}\ket{+}\ket{\mathcal{S}_j}  \ket{\bar{0}}& \nonumber \\
&\xrightarrow{~ U_{\mathcal H_l\cup \mathcal W_l}~ }& 
\begin{cases}
\ket{0} \ket{+} \ket{\mathcal{S}_j}  \ket{\bar{0}},  & ~\text{for} ~\mathcal{S}_j\notin (\mathcal H_l\cup\mathcal W_l) \\
\ket{1} \ket{+} \ket{\mathcal{S}_j}   \ket{\bar{0}},
   & ~\text{for} ~ \mathcal{S}_j \in  (\mathcal H_l \cup\mathcal W_l)
\end{cases}\nonumber\\
&\xrightarrow{~C-U_{Z}^{(m)}~ }& 
\begin{cases}
\ket{0} \ket{+} \ket{\mathcal{S}_j}  \ket{\bar{0}},  & ~\text{for} ~\mathcal{S}_j\notin (\mathcal H_l\cup \mathcal W_l) \\
\ket{1} \ket{+} \ket{\mathcal{S}_j}   \ket{Z_{\mathcal{S}_j}^{(m)}},
   & ~\text{for} ~ \mathcal{S}_j \in  (\mathcal H_l \cup \mathcal W_l)
\end{cases} 
\nonumber\\
& \xrightarrow{~\text{C-sign}~ } &
\begin{cases}
\frac{1}{\sqrt{2}} \left( \ket{0} \ket{0} \ket{\mathcal{S}_j}  \ket{\bar{0}}+ 
\ket{0} \ket{1} \ket{\mathcal{S}_j}  \ket{\bar{0}} \right),  & ~\text{for} ~\mathcal{S}_j\notin (\mathcal H_l\cup \mathcal W_l) \\
\frac{1}{\sqrt{2}} \left( \ket{1} \ket{0} \ket{\mathcal{S}_j}   \ket{Z_{\mathcal{S}_j}^{(m)}}+
\ket{1} \ket{1} \ket{\mathcal{S}_j}   \ket{-Z_{\mathcal{S}_j}^{(m)}}\right),   & ~\text{for} ~ \mathcal{S}_j \in  (\mathcal H_l \cup \mathcal W_l).
\end{cases} \nonumber\\
\end{eqnarray}
According to Lemma \ref{Lemma_Sparsitron_input1} and Lemma \ref{lemma_indicator_hl_wl}, it costs $\Ord{r\log n}$  and  $\Ord{r\log n \log\mu +\log^3\mu}$ to implement $C$-$U_{Z}^{(m)}$ and  $U_{\mathcal H_l\cup \mathcal W_l}$ run time respectively. Then the run time is  $\Ord{r\log\mu \log n+\log^3\mu +r\log n}= \Ord{r\log n\log \mu+ \log^3\mu } $ for implementing unitary $U_{in}^{(m)}$. 
\end{proof}

Treat the last register in Eq.~(\ref{eq_quantum_mrf_input}) as elements of a vector and the first  three registers as the index of the elements. By applying unitary $U_{\text{in}}^{(m)}$ to an initial state, we show that the resulting state can be used as the input of the quantum Sparsitron subroutine of Algorithm \ref{quantum_bounded-degree MRF_sparsitron}.
 
 \begin{lemma}\label{lemma_quantum_mrf_input}
Given the same hypothesis of Lemma \ref{lemma_sparsitron_input}, for each $m\in[M]$, let ${X}^{(m)}$ be the vector consisting of all $Z_{\mathcal{S}_j}^{(m)}$, where $\mathcal{S}_j \in( \mathcal H_l \cup \mathcal W_l)$.
  Starting from the following state
 \begin{equation}
\frac{1}{\sqrt{K}} \sum_{j_1=0}^n\cdots \sum_{j_{r-1}=0}^n \ket{0} \ket{+} \ket{j_1,\cdots,j_{r-1}} \ket{\bar{0}},
\end{equation}
where $K=(n+1)^{r-1}$, 
after applying a unitary $U_{in}^{(m)}$ defined in Lemma \ref{lemma_sparsitron_input}, we obtain a  quantum state 
\begin{eqnarray}
\frac{1}{\sqrt{2K}}\sum_{i=1}^{2K}\ket{g(i)}\ket{{\widetilde{X}}^{(m)}_i},
\label{eq_quantum_tilde_X}
\end{eqnarray}
which can be expressed as $\frac{1}{\sqrt{2K}}\sum_{i=1}^{2K}\ket{i}\ket{{\widetilde{X}}^{(m)}_{g^{-1}(i)}}$,
where ${\widetilde{X}}^{(m)}=(\bar{0},\bar{0},{X}^{(m)},-{X}^{(m)})\in \{0,1,-1\}^{2K} $ and ${\widetilde{X}}^{(m)}_i$ is the $i$-th element of ${\widetilde{X}}^{(m)}$, and $g(i)$ is a bijective map. 
 \end{lemma}
 \begin{proof}
 Starting from the following state
 \begin{equation}
\frac{1}{\sqrt{K}} \sum_{j_1=0}^n\cdots \sum_{j_{r-1}=0}^n \ket{0} \ket{+} \ket{j_1,\cdots,j_{r-1}} \ket{\bar{0}},
\end{equation} 
we apply a unitary $U_{in}^{(m)}$ defined in Lemma \ref{lemma_sparsitron_input}.  According to Eq.~(\ref{eq_quantum_mrf_input}), we have $\ket{0}\ket{+}\ket{\mathcal{S}_j}\ket{\bar{0}}$  for 
$ \mathcal{S}_j \notin (\mathcal H_l \cup \mathcal W_l)$ while $\frac{1}{\sqrt{2}}\ket{1}\ket{0}\ket{\mathcal{S}_j}\ket{Z_{ \mathcal{S}_j}^{(m)}}$ and $\frac{1}{\sqrt{2}}\ket{1}\ket{1}\ket{\mathcal{S}_j}\ket{-Z_{\mathcal{S}_j}^{(m)}}$ for $\mathcal{S}_j\in (\mathcal H_l \cup \mathcal W_l) $.    
Let ${X}^{(m)}$ be the vector consisting of all $Z_{I(\mathcal{S}_j)}^{(m)}=Z_{\mathcal{S}_j}^{(m)} $ where $\mathcal{S}_j \in(\mathcal H_l \cup \mathcal W_l)$. Then taking the first three registers as the index of the element in the last register, the result can be written as the state in Eq~(\ref{eq_quantum_tilde_X}). Notice that the number represented by the first three registers can be greater than $2K$ and they are not always continuous, because there are no terms such as $\ket{0}\ket{0}\ket{ \mathcal{S}_j}$ where $\mathcal{S}_j \in( \mathcal H_l \cup \mathcal W_l)$. There is a map from the index of elements and to the first three registers which we denoted as $g(i)$.  
In addition, we see that a zero padding is already added when $\mathcal{S}_j\notin(\mathcal H_l \cup \mathcal W_l)$.
\end{proof}

\subsection{Quantum Sparsitron algorithm setting and output}

Given quantum access to the training sets, there is a quantum Sparsitron algorithm that provides a polynomial speedup over the classical Sparsitron algorithm \ref{algosparsitron} in terms of the dimension of the samples \cite{rebentrost2021quantum}. The algorithm is shown in Algorithm \ref{algSparsitronQuantum} in the Appendix \ref{apendix_quantum_sparsitron}. 
The number of samples required and the run time of the quantum Sparsitron algorithm are given in the following theorem.
\begin{theorem}[Quantum Sparsitron \cite{rebentrost2021quantum}]\label{theoremQuantumSparsitron}
Let $\cal{D}$ be a distribution on $[-1,1]^n \times \{0,1\}$ where for $(X,Y) \sim \cal{D}$, $ \mathbb{ E}[Y|X = x] = \sigma (w \cdot x)$ for a non-decreasing 1-Lipschitz function $\sigma : \mathbbm R \to [0, 1]$. Suppose that $\Vert w\Vert_1\leq \lambda$ for a known $\lambda \geq 0$. Let $\epsilon,\rho \in (0, 1)$ and let there be given quantum access to $T+M = \Ord{\lambda^2(\log (n/\rho \epsilon))/ \epsilon^2}$ independent samples from $\mathcal D$.
Then, there is a quantum algorithm which returns $\left (t', {h^{(1)}},   \cdots, {h^{(t')}} , \Gamma^{(t')} \right)$, i.e, some $t'\in [T]$, inner product estimates, and a norm estimate. 
The run time of the algorithm to obtain this output is $\tOrd{{\lambda^2 T^2 M \sqrt{N}}\log \left(1/\rho \right)/{\epsilon } }$,
where $M = \Ord{ \log(T/\rho)/\epsilon^2}$.
Again, the algorithm can be run in an online manner.
Given this output, each coordinate $q_j$ of a vector $q \in \mathbbm R^N$ can be constructed separately in time $\Ord{T}$ and  $q$ satisfies  that
\be
\varepsilon(q) \leq \epsilon, 
\ee
where  $\varepsilon(q):= \mathbb{ E}_{(X,Y)\sim \mathcal{D}}\left[[\sigma(q\cdot X)-\sigma(w\cdot X)]^2\right]$ is the square loss function and
\begin{eqnarray}
q_j=\lambda  \frac {\beta^{\sum_{t=1}^{t'-1}\frac{1}{2}\left(1+\left(\sigma\left(\lambda h^{(t)}\right)-Y^{(t)}\right)X_j^{(t)}\right)}}{\Gamma^{(t')}}.
\end{eqnarray}
\end{theorem}

We now consider the quantum Sparsitron algorithm applied to the input defined in the previous section. 
We call the quantum Sparsitron Algorithm \ref{algSparsitronQuantum} with quantum access to $\widetilde{X}^{(t)}$ as Eq.~(\ref{eq_quantum_tilde_X}) and  $Y^{(t)}$, which can be obtained by $\left(1-Z_u^{(t)}\right)/2$, for $t\in [T]$.
The normalized weight vectors constructed by the algorithm are denoted by $\widetilde{q}_l^{(t)},\forall t\in[T]$, where the tilde denotes that these vectors have the same dimension as the input $\widetilde{X}^{(t)}$.
The algorithm uses the $2K$-dimensional initial weight vector $\widetilde{q}_l^{(1)}=\frac{1}{2K}(1,\cdots,1)$. For each round $t\in [T]$ and certain $t'\in [T]$, let  ${h}^{(t)} $ be the estimation of the inner product $\frac{\widetilde{q}_l^{(t)}}{ \Vert \widetilde{q}_l^{(t)} \Vert_1} \cdot \widetilde{X}^{(t)}$, and let ${\Gamma}^{(t')} $ be the estimation of the norm $ \left\Vert{\widetilde {q}_l^{(t')}} \right \Vert_1$ as line 3 and line 11, by Theorem \ref{theoremQuantumSparsitron} and the Algorithm \ref{algSparsitronQuantum}, we have
\be
\left \vert {h}^{(t)} - \frac{\widetilde{q}_l^{(t)}}{ \Vert \widetilde{q}_l^{(t)} \Vert_1} \cdot \widetilde{X}^{(t)} \right\vert &\leq& \frac{\epsilon}{32\lambda^2}, \nonumber\\
\left \vert{\Gamma}^{(t')} -  \left\Vert{ \widetilde{q}_l^{(t')}} \right \Vert_1 \right\vert &\leq& \left\Vert{ \widetilde{q}_l^{(t')}} \right \Vert_1 \frac{\epsilon}{32\lambda^2}\label{eq_h_Gamma}.
\ee
It is not necessary to know the map $g(i)$ in Eq.~(\ref{eq_quantum_tilde_X}), because each ${h}^{(t)}$ is estimated by using $1$-norm estimation as shown in Lemma $7$ in Ref.~\cite{rebentrost2021quantum}, and it is not affected by the index.   The quantum Sparsitron  Algorithm \ref{algSparsitronQuantum} with these inputs  then  outputs  $\left (t', {{h}^{(1)}},   \cdots, {{h}^{(t')}} , \Gamma^{(t')}\right)$. 
From this output, we can obtain  a quantum state representing the vector  $\widetilde{\bm{q}}_l=(\bm{q}_{l,1}, \bm{q}_{l,2},\bm{q}_{l,3},\bm{q}_{l,4})$ with dimension $2K$ and each element of which is non-negative. Then dimension of $\bm{q}_{l,3}$ and $\bm{q}_{l,4}$ are $\vert \mathcal H_l\vert +\vert \mathcal W_l\vert$ respectively while the dimension of $\bm{q}_{l,1}$ and $\bm{q}_{l,2}$ are $K-(\vert \mathcal H_l\vert +\vert \mathcal W_l\vert)$ respectively. 
Define the following distribution, which is the distribution used for applying Theorem \ref{theoremQuantumSparsitron}. 
\begin{defn}\label{definition_distribution_Dl}
Given the distribution $Z\sim \mathcal D$ on $\{-1,1\}^n$ a $r$-order multi-linear polynonial $p(Z)$ and $p_u =-2\partial_u p$ for $u\in[n]$.  For all $l\in[r-1]$,  let $\mathcal H_l$ and $\mathcal W_l(p_u)$ as in Definition \ref{SubsetH} and \ref{SubsetW}.
Define the distribution $\mathcal D'_l$ on $\{-1, 1\}^{\vert \mathcal H_l \cup \mathcal W_l \vert } \times \{0,1\}$ such that for $Z \sim \mathcal D$ we have $(X,Y) \sim \mathcal D'_l$ where $X=\left(Z_I :  I\in I(\mathcal H_l\cup \mathcal W_l)\right)$ and $Y=(1-Z_u)/2$.
\end{defn}
The run time of applying the Quantum Sparsitron is given in the following lemma. 
\begin{lemma}\label{lemma_quanutm_sparsitron_outcome}
Let $\mathcal{D}$ be a $r$-wise $\eta$-identifiable MRF on $Z\in \{-1,1\}^n$ as in Definition \ref{defMRFclass} and the degree of the underlying graph is bounded by $d$, and the  parameter $\mu= \min (2^d-1,d^{r-1})$.
Let $p_u$ as defined in Eq.~(\ref{definition_pu}) for a fixed $u\in [n]$ where $\Vert p_u\Vert_1 \leq 2\lambda$. 
Let $\mathcal H_l$, $\mathcal J_l(p_u)$, and $\mathcal W_l(p_u)$ be sets as in Definitions \ref{SubsetH}, \ref{SubsetJ}, and \ref{SubsetW}, defined with respect to the value $\eta$.
Let $\bm q_{l}= (\widehat q_l(I) : I\in I(\mathcal H_l\cup \mathcal W_l))$,  $\bm p_{u,l}= (\widehat p_u(I) : I\in I(\mathcal H_l\cup \mathcal W_l))$, and $(X,Y) \sim \mathcal{D}'_l$ as in Definition \ref{definition_distribution_Dl}.
Given quantum access to $T+M$ samples as Definition \ref{MRF_input}, $\epsilon >0,  \rho \in (0,1)$, calling the Quantum Sparsitron for each $l\in [r-1]$ with inputs 
 $$\left(\epsilon,  \frac{\rho}{2n(r-1)}, 2\lambda, (\widetilde{X}^{(i)}, Y^{(i)})_{i=1}^{T+M},  T, M\right),$$
with the number of samples
\begin{eqnarray}
T &\in& \Ord{\lambda^2(\log(rn^r/\rho \epsilon))/\epsilon^2}, \nonumber\\
M &\in& \Ord{ \log(T/\rho)/\epsilon^2},\nonumber
\end{eqnarray}
we can obtain the
values $\mathcal V := \left (t', {{h}^{(1)}},   \cdots, {{h}^{(t')}}, {\Gamma}^{(t')}\right)$  in time $\tOrd{\mu +\frac{( \lambda {T})^2 M \sqrt{n^{r-1}}}{\epsilon }\log \frac{1}{\rho} }$, with success probability at least $1-\frac{\rho}{2n(r-1)}$, where $t'\in[T]$ and the values $\mathcal V$ satisfy the guarantees defined in Eq.~(\ref{eq_h_Gamma}).
From $\mathcal V$, we can construct a vector $\bm q_l$ which is the coefficient vector of polynomial $q_l$ such that 
\be
 \mathbb{ E}_{(X,Y)\sim \mathcal{D}'_l}[\left(\sigma({ q_l})-\sigma({{p_{u,l}}})\right)^2] \leq \epsilon.
\ee
\end{lemma}

\begin{proof}
According to Eq~(\ref{equ_mrf}), we can see that the distribution $\mathcal D'_l$ satisfies
\begin{eqnarray}
\mathbb{ E}_{(X,Y) \sim \mathcal D'_l}\left[Y | X \right] = \sigma( \bm{p}_{u,l} \cdot X). 
\end{eqnarray}
In addition, by assumption is holds that
$\Vert p_u\Vert \leq 2\lambda$.
Choose the integers $T,M$ such that
\begin{eqnarray}
T & =& {C \lambda^2\log(rn^r/\rho \epsilon)}/{\epsilon^2}, \nonumber\\
M &= &{ C \log(T/\rho)}/{\epsilon^2},\nonumber
\end{eqnarray}
with a sufficiently large constant $C>0$.
Consider the input preparation for all $i \in[T+M]$. 
We are given quantum samples from Definition \ref{MRF_input}. Let ${\widetilde{X}}^{(i)}=\left(\bar{0},\bar{0},{X}^{(i)},-{X}^{(i)}\right)\in \{0,1,-1\}^{2K} $,  $ K={(n+1)^{r-1}}$  as in Lemma \ref{lemma_quantum_mrf_input} and   $Y^{(i)}=\left(1-Z_u^{(i)}\right)/2.$
By Lemma \ref{Lemma_Sparsitron_input1}, Lemma \ref{lemma_indicator_hl_wl} and Lemma \ref{lemma_sparsitron_input},  it requires at most $C_1=\tOrd{\mu \log n}$ time to construct a sorted QRAM for $\mathcal W_l$ and $C_2 = \Ord{r\log n\log \mu +\log^3\mu}$ time to prepare the input $\left(\widetilde{{X}}^{(i)},Y^{(i)}\right)\in \{0,1,-1\}^{2K} \times \{0,1\}$.
Hence, we obtain quantum access of $\left(\widetilde{{X}}^{(i)},Y^{(i)}\right)$ for all $i\in [M+T]$.

Now we have the complete hypothesis for Theorem \ref{theoremQuantumSparsitron} and obtain the corresponding guarantees and run time. First, consider the run time.
According to Theorem \ref{theoremQuantumSparsitron}, the total run-time is  
\be
& & \tOrd{C_1+\frac{C_2 \lambda^2 {T}^2 M \sqrt{2(n+1)^{r-1}}}{\epsilon }\log \frac{1}{\rho} }\nonumber \\
& = &\tOrd{\mu\log n+\frac{\left(r\log n\log \mu + \log^3\mu \right) \lambda^2 {T}^2 M \sqrt{2(n+1)^{r-1}}}{\epsilon }\log \frac{1}{\rho} } \nonumber \\
& = & \tOrd{\mu+\frac{\lambda^2 {T}^2 M \sqrt{n^{r-1}}}{\epsilon }\log \frac{1}{\rho}}
\ee
Next, we discuss the guarantee. The algorithm outputs the values $\left( t', {{h}^{(1)}},   \cdots, {{h}^{(t')}}, {\Gamma}^{(t')}\right)$.
From Theorem \ref{theoremQuantumSparsitron}, there exists a vector  $\widetilde{\bm q_l}=\left(\bm q_{l,1},\bm q_{l,2},\bm q_{l,3},\bm q_{l,4}\right) \in \mathbbm R^{2K}$ such that each single element $\widetilde{q}_{l,j}$ can be constructed separately via
\begin{eqnarray}
\widetilde{q}_{l,j}=2\lambda  \frac {\beta^{\sum_{t=1}^{t'-1}\frac{1}{2}\left(1+\left(\sigma(2\lambda {h}^{(t)})-Y^{(t)}\right)\widetilde{X}_j^{(t)}\right)}}{{\Gamma}^{(t')}}. \label{eq_tilde_q}
\end{eqnarray}
We obtain our output polynomial $q_l$ with coefficient vector $ \bm q_l= \bm q_{l,3}- \bm q_{l,4}$. According to Eq. (\ref{eq_tilde_q}), each element of $\bm q_{l}$ is obtained  by calculating
\begin{eqnarray}
{q}_{l,j}=2\lambda  \frac {\phi \left({X}_j^{(1)},\cdots, {X}_j^{(t')} \right) - \phi \left(-{X}_j^{(1)},\cdots, -{X}_j^{(t')} \right) }{{\Gamma}^{(t')}},\label{eq_q}
\end{eqnarray}
where $\phi \left({X}_j^{(1)},\cdots, {X}_j^{(t')} \right)= \beta^{\sum_{t=1}^{t'-1}\frac{1}{2} \left(1+\left(\sigma(2\lambda {h}^{(t)})-Y^{(t)}\right){X}_j^{(t)}\right)}$. By Theorem \ref{theoremQuantumSparsitron} and using the fact that $\widetilde{\bm q}_l\cdot \widetilde{ X}={\bm q}_l\cdot{X}$, the square loss function satisfies
\be
\mathbb{ E}_{(X,Y)\sim \mathcal{D}'_l}\left[\left (\sigma({ q_l})-\sigma({ p_{u,l}})\right )^2 \right] \leq \epsilon.
\ee
Hence, we obtain a guarantee for learning $\bm p_{u,l}$.
\end{proof}
We construct a unitary that prepares quantum state which is a superposition of the vector $\bm q_l$. We describe the procedure to construct the quantum state in Eq.~(\ref{eq_vector_q})
in the following. 
\begin{lemma}\label{lemma_quantum_element_preparation}
In the same setting as Lemma \ref{lemma_quanutm_sparsitron_outcome},
let $\mathcal V$ and $\bm q_l$ be the output of Lemma \ref{lemma_quanutm_sparsitron_outcome}.
Let $K=(n+1)^{r-1}$.
Given the unitary $U_{\mathcal F_l}$ in Lemma \ref{lemma_indicator_subsets_Fl} for the string set $\mathcal F_l(p_u)$ defined in Definition \ref{SubsetF}, 
a unitary $U_q$ can be constructed which prepares the following quantum state 
\begin{eqnarray}
 \frac{1}{\sqrt{2K}}\left( \sum_{\mathcal{S}_j \in \mathcal F_l}\ket{1} \ket{\mathcal{S}_j} \ket{\widehat{q}_l(I(\mathcal{S}_j)) }+ \sum_{\mathcal{S}_j \notin \mathcal F_l}\ket{0}  \ket{\mathcal{S}_j} \ket{\bar{0}}\right),\label{eq_vector_q}
\end{eqnarray}
in time  $\Ord{T\left( r\log n\log \mu+ \log^3\mu \right)}$. 
\end{lemma} 
\begin{proof}
As discussed before, strings in $\mathcal F_l$  correspond to potential maximum monomials containing $l$ variables of polynomial $p_u$. We first select strings in $\mathcal F_l$ by using Lemma \ref{lemma_indicator_subsets_Fl}. Then  calculate each $q_l(I(\mathcal{S}_j))$ for strings $\mathcal{S}_j$ in $\mathcal F_l$ by using Eq.~(\ref{eq_q}). 

The procedure of constructing the state in Eq.~(\ref{eq_vector_q}) is shown in the following: 
\begin{enumerate}[(\romannumeral1)]
\item Start from state  $\frac{1}{\sqrt{K}}\sum_{\mathcal{S}_j}\ket{0}\ket{\mathcal{S}_j}\ket{\bm 0}$ with $7$ registers for all $\mathcal{S}_j=(j_1,\cdots j_{r-1}) \in ([n] \cup \{0\})^{r-1}$, where  $\ket{\bm 0}=\ket{\bar{0}}\ket{\bar{0}}\ket{\bar{0}}\ket{\bar{0}}\ket{\bar{0}}$. 
\item  Apply $U_{\mathcal F_l}$ defined in Lemma \ref{lemma_indicator_subsets_Fl} on the two registers, we obtain
$$\sum_{\mathcal{S}_j \in \mathcal F_l}\ket{1} \ket{\mathcal{S}_j}\ket{\bm 0} + \ket{\chi},$$ where $\ket{\chi}= \sum_{\mathcal{S}_j \notin \mathcal F_l} \ket{0}  \ket{\mathcal{S}_j}\ket{\bm{0}}$, and for simplicity, we ignore the normalization factor $\sqrt{1/2K}$ for the moment.
\item For each $t\in [t']$, apply controlled $U_{Z}^{(t)}$  as in Lemma \ref{Lemma_Sparsitron_input1} on the first three registers, controlled by the first register in state $\ket{1}$
$$ \sum_{\mathcal{S}_j \in \mathcal F_l}\ket{1} \ket{\mathcal{S}_j}\ket{ Z_{\mathcal{S}_j}^{(1)},\cdots,  Z_{\mathcal{S}_j}^{(t')}}\ket{\bar{0}} \ket{\bar{0}} \ket{\bar{0}} \ket{\bar{0}} + \ket{\chi}.$$
\item Let $\ket{Z'_{\mathcal{S}_j}}:=\ket{ Z_{\mathcal{S}_j}^{(1)},\cdots,  Z_{\mathcal{S}_j}^{(t')}}$.  Apply CNOT gates to copy each $Z_{\mathcal{S}_j}^{(t)} $ in the third register to the fourth register and followed by a $X$ gate on the sign qubit of the fourth register of each $Z_{\mathcal{S}_j}^{(t)}$, it yields
  \begin{eqnarray}
 \sum_{\mathcal{S}_j \in \mathcal F_l}\ket{1} \ket{\mathcal{S}_j}\ket{Z'_{\mathcal{S}_j}}\ket{-Z'_{\mathcal{S}_j}}\ket{\bar{0}}\ket{\bar{0}}\ket{\bar{0}} + \ket{\chi}.\nonumber
      \end{eqnarray}
\item Combining with the values ${h}^{(t)}$ and $Y^{(t)}$ for all $t\in[t']$, and ${\Gamma}^{(t')} $ and  calculating  $\widetilde{q}_{l,j}$ according to Eq.~$(\ref{eq_tilde_q})$, where each $Z_{\mathcal{S}_j}^{(t)}$ and $-Z_{\mathcal{S}_j}^{(t)}$ is one element of $\widetilde{X}^{(t)}$,   we have 
    \begin{eqnarray}
& & \sum_{\mathcal{S}_j \in \mathcal F_l}\ket{1} \ket{\mathcal{S}_j}\ket{Z'_{\mathcal{S}_j}}\ket{-Z'_{\mathcal{S}_j}}\ket{q_{l,3}(I(\mathcal{S}_j))}\ket{q_{l,4}(I(\mathcal{S}_j))} \ket{\bar{0}} + \ket{\chi}. \nonumber
\end{eqnarray}
 \item Calculate $\widehat{q}_l(I(\mathcal{S}_j))=q_{l,3}(I(\mathcal{S}_j))-q_{l,4}(I(\mathcal{S}_j))$ and store it in the last register,
 \begin{eqnarray}
 \sum_{\mathcal{S}_j \in \mathcal F_l}\ket{1} \ket{\mathcal{S}_j}\ket{Z'_{\mathcal{S}_j}}\ket{-Z'_{\mathcal{S}_j}}\ket{q_{l,3}(I(\mathcal{S}_j))}\ket{q_{l,4}(I(\mathcal{S}_j))} \ket{\widehat{q}_l(I(\mathcal{S}_j)) } + \ket{\chi}.\nonumber
\end{eqnarray}
\item Undo steps \romannumeral5,\romannumeral4,\romannumeral3, it yields state (here including the normalization factor again)
\begin{eqnarray}
 \frac{1}{\sqrt{2K}}\left ( \sum_{\mathcal{S}_j \in \mathcal F_l}\ket{1} \ket{\mathcal{S}_j}\ket{\bar{0}}\ket{\bar{0}}\ket{\bar{0}}\ket{\bar{0}} \ket{\widehat{q}_l(I(\mathcal{S}_j)) }+ \ket{\chi} \right).
\end{eqnarray}
\end{enumerate}
We have disentangled the third up to the sixth register, hence we obtain the state in Eq.(\ref{eq_vector_q}). 

For the time complexity, according to Lemma \ref{lemma_indicator_subsets_Fl}, step (\romannumeral2) of the above procedure takes time $\tOrd{\kappa \log n}$ to construct the sorted QRAM for $\bar{\mathcal F}_l$, and time $\Ord{r\log \kappa\log n+\log^3\kappa}$ to  implement the unitary $U_{\mathcal F_l}$. For step (\romannumeral3), by Lemma \ref{lemma_sparsitron_input}, applying $U_Z^{(t)}$ and $U_Y^{(t)}$ for $t'$ samples costs time $\Ord{t'\left(r\log n\log\mu+ \log^3\mu \right)}$ which is bounded by $\Ord{T\left( r\log n\log\mu+ \log^3\mu \right)}$. For step (\romannumeral4), the number of CNOT and X gates required are at most $T$ respectively. 
For step (\romannumeral5) and (\romannumeral6), as the summation in Eq.~(\ref{eq_tilde_q}) requires run time at most $T$, these steps can be implemented in time $\Ord{T}$.
Therefore, the total run time for the procedure is then $\Ord{T\left( r\log n\log\mu+ \log^3\mu\right)}$.
\end{proof}

\subsection{Quantum search for maximal monomials}

We now show that we can find all maximal monomials with size $l$ of $p_u$.
According Lemma  \ref{lemma_quanutm_sparsitron_outcome}, we can construct a quantum state of vector $\bm{q}_l$.  
As the square loss function satisfies
\be
\mathbb{ E}_{(X,Y)\sim \mathcal{D'}}[(\sigma({q_l})-\sigma( p_{u,l}))^2] \leq \epsilon,
\ee
from Lemma \ref{lemma_threshold_eta1} in the classical algorithm part, we can see whether $I(\mathcal{S}_j)$ is a maximal monomial or not by the value of $\widehat{q}_l(I(\mathcal{S}_j))$. 
By using a quantum search algorithm,  we can find all subsets with size $l$ for which the corresponding coefficient is greater than $\eta$.
Thus, each subset found in this step is one of the maximal monomials of $p_u$. The run time of this step is given in the following lemma. 

\begin{lemma}\label{lemma_quantum_subset_finding}
In the same setting for an MRF as Lemma \ref{lemma_quanutm_sparsitron_outcome}, and given the unitary $U_q$ from Lemma \ref{lemma_quantum_element_preparation}.
Let $\rho \in (0,1)$,  $\eta>0$.
Then,  we can find all maximal monomials with size $l$ of $p_u$ in time $\tOrd{\kappa + T\sqrt{\mu n^{r-1}}\log \left(1/\rho \right) }$ with success probability $1-\frac{\rho}{2n(r-1)}$, where $\kappa =\min (2^d/\sqrt{d}, d^{r-2})$.
\end{lemma}
\begin{proof}
As proved in Lemma \ref{lemma_quantum_element_preparation}, the  quantum state in Eq.~(\ref{eq_vector_q})
can be prepared in time
\begin{eqnarray}\label{eq_state_prepare_time}
T_p \in  \Ord{T\left( r\log n\log\mu+ \log^3\mu\right)}.
\end{eqnarray}
In the following, we will show that we can find all strings in set $\mathcal F_l$ which the coefficient of the corresponding subset is $\geq\eta$ by using quantum counting (Lemma \ref{lemma_quantum_counting}) and quantum search (Lemma \ref{Tight_bound_quanutm_search}).  If there exists $k$ such strings, by using Lemma \ref{lemma_quantum_counting}, we can find $k'$  in time  
\be
T_{\text{qcount}} \in \Ord{T_p\sqrt{kn^{r-1}}\log \frac{1}{\rho_\text{qc}}},
\ee
such that $k'=k$ with probability $1-\rho_\text{qc}$, for $\rho_\text{qc} \in (0,1)$.  If $k=0$, we can determine it in time $\Ord{\sqrt{n^{r-1}}}.$ 
After obtaining $k$, we then use quantum search algorithm as Lemma \ref{Tight_bound_quanutm_search} to get such a string.  According to Lemma \ref{Tight_bound_quanutm_search} and the time $T_p$ in  Eq.~(\ref{eq_state_prepare_time})
to prepare the state (\ref{eq_vector_q}), if there are $k$ ($k>0$) solutions in  $2(n+1)^{r-1}$ states, with success probability $1-\rho_\text{qs}$ ($\rho_\text{qs} \in(0,\frac{1}{2})$),   we can  find one solution in time  
\be
T_\text{qs} \in \Ord{T_p\sqrt{\frac{n^{r-1}}{k}}\log \frac{1}{\rho_\text{qs}}},
\ee
as $2(n+1)^{r-1}$ is bounded by $\Ord{n^{r-1}}.$    
 
 For each round of Grover search, if we get a $\widehat{q}_l(I(\mathcal{S}))$ is less than $\eta$, we know that we fail, then we do another Grover search.  As discussed before, we can find one solution in time $T_\text{qs}$ with success probability  ${1-\rho_\text{qs}}$.  Finding all $k$ solutions can be related to a coupon collection problem. Since the probability to find the $i$-th new coupon in $k$ coupons is $\rho_i=\frac{k-i+1}{k}$,   the expected run time to obtain all $k$ different solutions is then given by
$\mathbb E \left[\widetilde{T}\right]= T_\text{qs}\sum_{i=1}^k \frac{1}{\left(1-\rho_\text{qs}\right) \rho_i} = T_\text{qs} \frac{1}{1-\rho_\text{qs}}k \left(1+\frac{1}{2}+\cdots + \frac{1}{k}\right).$  Choosing $\rho_\text{qs}$ as a constant which is less than $\frac{1}{2}$,  we have
 \begin{eqnarray}
\mathbb E \left[\widetilde{T}\right]\in\Ord{T_{g} k \log k}=\Ord{T_p\sqrt{\frac{n^{r-1}}{k}} k \log k}.
 \end{eqnarray}
Here, we use the fact that $k \left(1+\frac{1}{2}+\cdots + \frac{1}{k}\right)$ is bounded by $\Ord{k\log k}.$
  By Markov's inequality, the problem can be solved in time $2 \mathbb E \left[\widetilde{T}\right]$ with probability greater than $1/2$. Given $\rho_{\text{cp}} \in(0,1)$, repeating $\log \frac{1}{\rho_{\text{cp}}}$ times, the probability that all the repetitions cost time longer than $2\mathbb E \left[\widetilde{T}\right]$ is less than $(\frac{1}{2})^{\log\frac{1}{\rho_{\text{cp}}} }={\rho_{\text{cp}}}$.  
The success probability of obtaining all solutions is bounded by $1-{\rho_{\text{cp}}}$. The run time is then  
\be
T_{\text{cp}} \in \Ord{T_p \sqrt{\frac{n^{r-1}}{k}} {k \log k}\log \frac{1}{\rho_{\text{cp}}}  }\subseteq\tOrd{T_p \sqrt{kn^{r-1}} \log \frac{1}{\rho_{\text{cp}}}  } .
\ee
Let $\rho_{\text{cp}}=\rho_\text{qc}=\frac{\rho}{4n(r-1)}$, by union bound, the total success probability is then bounded by $1-\rho_\text{qc}-\rho_{\text{cp}}=1-\frac{\rho}{2n(r-1)}$.
Given the QRAM for $\bar{\mathcal F}_l$,  the value $k$ and all the $k$ solutions can be found in time 
\be
T_{\text{qcount}}+T_{\text{cp}} & \in &  \tOrd{T_p\sqrt{kn^{r-1}}\left(\log \frac{1}{\rho_\text{qc}}+ \log \frac{1}{\rho_\text{cp}}\right)}\nonumber\\
&=& \tOrd{T \left( r\log n\log\mu+ \log^3\mu\right)\sqrt{kn^{r-1}}\log \frac{1}{\rho} }.
\ee
As $k$ is bounded by $\mu$, the run time is then $ \tOrd{ T\sqrt{\mu n^{r-1}}\log \frac{1 }{\rho})}$ ($r$ is bounded by $\Ord{\log \mu}$). 
Combining with the cost of constructing the QRAM for $\bar{\mathcal F}_l$, the total run time of Line \ref{quantum_mrf_step_search} in Algorithm \ref{quantum_bounded-degree MRF_sparsitron} is then given by 
\be\tOrd{ \kappa \log n + T \sqrt{\mu n^{r-1}}\log ({1}/{\rho})}=\tOrd{ \kappa + T \sqrt{\mu n^{r-1}}\log (1/\rho)}.
\ee

\end{proof}

\subsection{Main result for quantum MRF structure learning algorithm}

Now we present the quantum MRF structure learning algorithm which is Algorithm \ref{quantum_bounded-degree MRF_sparsitron}. 
As the classical algorithm, for a node $u$, we start from $l=r-1$, by using the quantum Sparsitron algorithm and quantum search with quantum access to  the samples, we can find all the maximal monomials containing $l$ variables of $p_u$. By decreasing $l$,  we can find all maximal monomials of the polynomial $p_u$ with a high success probability. 
Based on the results, we can find all neighbors of node $u$. We can recover the structure of the underlying graph by running the algorithm for each node.

 \begin{figure}[h!]
\begin{algorithm}[H]
  \caption{Quantum Bounded-degree MRF structure learning via Sparsitron (\textsc{Qmrf}) }\label{quantum_bounded-degree MRF_sparsitron}
  \begin{algorithmic}[1] 
  \Require{Quantum access to $T+M$ samples on $Z\in \{1,-1\}^n$, given parameter $\lambda >0$,  probability $\rho \in(0,1)$,
  $\eta >0$, $u\in[n]$, $r \in [n]$, $d\in [n-1]$. 
     }
    \State Initialize $J, S \leftarrow \emptyset$,   subset $I\subset [n]\setminus \{u\}$,  and $l \gets r-1$. 
    \State $K\leftarrow (n+1)^{r-1}$.
    \While{$|S|<d$ and $l \neq 0$}
    \State $J'_l\gets J$.
    \State  $W'_l \leftarrow  \{I \mid \exists A \in J'_l, I \subseteq A \wedge \vert I \vert >l \}$. 
\State For $i\in[T+M]$,  construct quantum query access for vector  $\widetilde{X}^{(i)}=\left(\bar{0},\bar{0},{X}^{(m)},-{X}^{(m)}\right)\in \{0,1,-1\}^{2K}$ as in Lemma \ref{lemma_quantum_mrf_input} and $ Y^{(i)}=\left(1-Z_u^{(i)} \right)/2 \in \{0,1\}$, ${X}^{(i)}$ is the vector consisting of products $Z_I^{(i)}=\prod_{k\in I}Z_k^{(i)}$ for all subsets $I \in W'_l$ and all subsets $I$ where $\vert I \vert \leq l$.  
\State $\epsilon \leftarrow e^{\Ord{\lambda}} e^{\Ord{\lambda r}}  \eta^2$
     \State\label{quantum_mrf_step_sparsitron}  $\left (t', {{h}^{(1)}},   \cdots, {{h}^{(t')}} , {\Gamma}^{(t')}\right) \leftarrow$  Apply the \textsc{Quantum\_Sparsitron} Algorithm
      \ref{algSparsitronQuantum} with input  $$\left(T,M,\epsilon,  \frac{\rho}{2n(r-1)}, 2\lambda, \left(\widetilde{X}^{(i)}, Y^{(i)}\right)_{i=1}^{T+M}\right),$$ by Lemma \ref{lemma_quanutm_sparsitron_outcome}. 
        \State\label{quantum_mrf_step_search} Find all subsets $I$ with size $\vert I\vert =l$ and the coefficient satisfies $\vert \widehat q_l(I)\vert \geq{\eta}$  with success probability $1-\frac{\rho}{2 n(r-1)}$ by Lemma \ref{lemma_quantum_subset_finding}. 
               \For {each  $I$ found }
         \label{algorithm_quantum_rbm_for}
     \State 
     $J \gets J \cup \{I\}$, $S\gets S\cup I$.
 \EndFor
    \State $l \gets l-1$.
        \EndWhile 
 \Ensure{S}
  \end{algorithmic}
\end{algorithm}
\end{figure}

We give the required number of samples, run time, and success probability of Algorithm \ref{quantum_bounded-degree MRF_sparsitron} in the following theorem.

\begin{theorem}[Quantum MRF structure learning]\label{theorem_quantum_MRf_structure_learning} 
Let $\mathcal{D}$ be an $\eta$-identifiable MRF on $Z \in \{-1,1\}^n$ with $r$-order interactions and the degree of the underlying graph $G$ is bounded by $d$, the factorization polynomial of the MRF $p(Z)=\sum_{I\in C_r(G)}\widehat{p}(I)Z_I$ with  $ \max_i\Vert \partial_ip\Vert \leq \lambda$.  Given $T+M$ samples from the MRF, $\rho \in(0,1/2)$
when the number of samples satisfies
$$T+M \in {e^{\Ord{r}} e^{\Ord{ \lambda r}}} \log (nr/\rho \eta)/\eta^4, $$
 the structure of the underlying graph can be recovered  in time $$\tOrd{ n\mu  + \left( {  {T} M }/{\eta^2 }  +  \sqrt{\mu} \right) T \sqrt{n^{r+1}} \log ({1}/{\rho}) }$$ with success probability $1-\rho$,   by performing Algorithm \ref{quantum_bounded-degree MRF_sparsitron} for $n$ vertices,  where $T \in e^{\Ord{r}} e^{\Ord{ \lambda r}} \log (nr/\rho \eta)/\eta^4$ and $M \in e^{\Ord{r}} e^{\Ord{\lambda r}} \log( T/\rho)/\eta^4$ and $\mu =\min (2^d-1, d^{r-1}).$
\end{theorem}

\begin{proof}
As proved in Theorem \ref{th:identifiable} for the classical algorithm \ref{alg_classical_MRF_sparsitron}. For each $l\in [r-1]$, we have $I(\mathcal{J}_l(p_u))=J'_l$, $I(\mathcal{W}_l(p_u))=W'_l$ for $J'_l$ and $W'_l$ in Algorithm \ref{quantum_bounded-degree MRF_sparsitron}. 

We now show the number of samples required. By Lemma \ref{lemma_quanutm_sparsitron_outcome}, the number of samples needed for each call to quantum Sparsitron algorithm in Line \ref{quantum_mrf_step_sparsitron} of  Algorithm \ref{quantum_bounded-degree MRF_sparsitron}
\begin{eqnarray}
T+M & \in &\Ord{\lambda^2\log(rn^r/\rho \epsilon)/\epsilon^2} \subseteq e^{\Ord{r}}  e^{\Ord{\lambda r}}\log(nr/\rho \eta)/\eta ^4,\nonumber\\
T & \in & e^{\Ord{r}}  e^{\Ord{\lambda r}} \log(nr/\rho \eta) / \eta ^4,\nonumber\\
 M & \in & e^{\Ord{r}} e^{\Ord{\lambda r}}\log( T/\rho)/\eta^4,
\end{eqnarray}
where the results are obtained by using  $\epsilon = c e^{-6 -4\lambda -2\lambda(r-1)} \eta^2/(2^{r+1})$ for a small constant $ c>0 $ by Lemma \ref{lemma_threshold_eta0}.

We analyze the run time. The most time-consuming steps are Line \ref{quantum_mrf_step_sparsitron} and Line \ref{quantum_mrf_step_search}.  
By Lemma \ref{lemma_quanutm_sparsitron_outcome}, the run time of performing the quantum Sparsitron algorithm of Line \ref{quantum_mrf_step_sparsitron} of Algorithm \ref{quantum_bounded-degree MRF_sparsitron}  is  $\tOrd{\mu +\frac{(\lambda {T})^2 M \sqrt{n^{r-1}}}{\epsilon }\log \frac{1}{\rho} }.$  From 
Lemma \ref{lemma_quantum_subset_finding}, the Line \ref{quantum_mrf_step_search}  takes a run time of $\tOrd{\kappa + T\sqrt{\mu n^{r-1}}\log \frac{1}{\rho}  }.$   For each step of the loop of Algorithm \ref{quantum_bounded-degree MRF_sparsitron}, the total run time of Line \ref{quantum_mrf_step_sparsitron} and Line \ref{quantum_mrf_step_search} is given by
\begin{eqnarray}
 T_{\rm inner} &\in& \tOrd{\mu  +\frac{(\lambda {T})^2 M \sqrt{n^{r-1}}}{\epsilon }\log \frac{1}{\rho} +\kappa + T\sqrt{\mu n^{r-1}} \log\frac{1}{\rho}  }\nonumber\\
&=& \tOrd{ \mu  +\left( \frac{ \lambda^2 {T} M }{\epsilon }  +  \sqrt{\mu} \right) T \sqrt{n^{r-1}} \log \frac{1}{\rho} }. 
\end{eqnarray}
With at most $r-1$ runs for the \emph{while} loop, the run time becomes  $\tOrd{ rT_{\rm inner} }$. Hence, for $n$ vertices, the total run time is  $\tOrd{ nr T_{\rm inner} }$. 

Notice that $T$ already contains the factor $e^{\Ord{r}}$ and $e^{\Ord{\lambda r}}$, $r \leq \log \mu$,  the run time can be written as $\tOrd{ n T_{\rm inner} }.$  As we have  $\epsilon = c\ e^{-6 -4\lambda -2\lambda(r-1)} \eta^2/(2^{r+1})$ with a small constant $c>0$, the run time can be written as    
 \begin{eqnarray} 
& &\tOrd{ n\mu  + \left( \frac{ \lambda^2 {T} M e^{2\lambda} e^{2r\lambda} 2^r  }{\eta^2 }  +  \sqrt{\mu}  \right) T \sqrt{n^{r+1}} \log \frac{1}{\rho} }\nonumber\\
&=& \tOrd{ n\mu  + \left( \frac{ {T} M  }{\eta^2 }  +  \sqrt{\mu} \right) T \sqrt{n^{r+1}} \log \frac{1}{\rho} },
 \end{eqnarray}
because the factors $\lambda^2e^{2\lambda}e^{2r\lambda2^r}$ are contained in $T$. 

For the success probability, as the success probability of Line \ref{quantum_mrf_step_sparsitron} and Line \ref{quantum_mrf_step_search} are at least $1-\frac{\rho}{2n(r-1)}$ respectively,  by using Boole's inequality, the total success probability of Algorithm  \ref{quantum_bounded-degree MRF_sparsitron}  is at least  $1-\frac{2\rho}{2n(r-1)}({r-1})= 1-\frac{\rho}{n} $. Then the probability of success for $n$ vertices bounded by $  1-\rho.$
\end{proof}
\section{Discussion and conclusion}
 \label{section_conclusion}
We have presented a modified classical structure learning algorithm for $r$-wise and $\eta$-identifiable MRFs. The time complexity is the same as the original algorithm in Ref.~\cite{klivans2017learning}.  Based on the classical algorithm, we have developed a quantum structure learning  algorithm for MRFs with the same parameters $r,\eta$, and bounded degree. We have proved that the quantum structure learning algorithm  provides polynomial speedup over the corresponding classical algorithm, in terms of the dimension of the MRF. The number of training samples required is the same as the classical algorithm, which is logarithmic in the dimension of the samples.

From Theorem \ref{th:identifiable}, the run time of our classical algorithm can be written as
\be
\tOrd{ n^r M},
\ee
where $M=Ce^{\Ord{r}} e^{\Ord{ \lambda r}} \log \left (1/{\rho} \right) /\eta^4$ for a constant $C>0$. 
From Theorem \ref{theorem_quantum_MRf_structure_learning}, the run time of the quantum algorithm can be written as
\be
\tOrd{ n\mu  + \left( \frac{\log^2\left(1/\rho\right) }{\eta^{10} }  +  \sqrt{\mu} \right) \sqrt{n^{r+1}} \times M \log\frac{1}{\rho} }.
\ee
Assume an MRF with degree $d\in \Ord{\log n}$, then  $\mu \in \Ord{n}$. In addition, assume that $\eta<1$ and $r>2$.
The quantum run time simplifies to 
\be
& &\tOrd{ n^2  + \left( \frac{\log^2\left(1/\rho\right) }{\eta^{10} }  +  \sqrt{n} \right) \sqrt{n^{r+1}} \times  M \log \frac{1}{\rho} }\\
& &\subseteq  
\tOrd{\frac{\sqrt{n}  \log^3\left(1/\rho\right)} {\eta^{10}} \times \sqrt{n^{r+1}} \times M}.
\ee
Hence, for such MRFs, the quantum algorithm provides a speedup over the classical counterpart if
\be
 C_2\sqrt[r-2]{\frac{\log^6\left(1/\rho\right)} {\eta^{20}} } < n,
\ee
for a constant $C_2>0$. Similarly, for MRFs with  $\eta>1$ and $r>2$, the quantum algorithm is more efficient if
\be
 C_3 \sqrt[r-2]{\log^6\left(1/\rho\right)} < n,
\ee
for a constant $C_3>0$.

The run time of the classical algorithm is nearly optimal, given the hardness of learning $r$-wise Markov random fields discussed in Appendix A of Ref.~\cite{klivans2017learning}.
It was proved via a reduction to learning sparse parities with noise (LSPN) that learning $k+1$-MRFs is harder than $k$-LSPN. 
The best-known algorithm for learning $k$-LSPN is due to Valiant \cite{valiant2015finding} and runs in time $n^{\Omega(0.8k)}$. 
As discussed above, for an MRF with degree $d\in \Ord{\log n}$, in terms of the dimension, the run time of the quantum algorithm  is $ \tOrd{n^{r/2+1}}$. We see that it is  smaller than the known classical lower bound 
$n^{ \Omega(0.8(r-1))}$.

In addition, inspired by binary search, we have constructed a scheme for quantum set membership queries with poly-logarithmic run time. This scheme can be useful for constructing other efficient quantum algorithms.

\section{Acknowledgements}

We acknowledge valuable discussions with Miklos Santha. 
This work was supported by the Singapore National Research Foundation, the Prime Minister’s Office, Singapore, the Ministry of Education, Singapore under the Research Centres of Excellence programme under research grant R 710-000-012-135.

\bibliographystyle{apsrev}
\bibliography{MRF}

\appendix 
\section{Classical Sparsitron Algorithm and MRF Lemmas}
\label{app_classical_sparsitron}

The classical Sparsitron  algorithm is shown in  Algorithm \ref{algosparsitron}, where the learning rate $\beta$ can be set as $\beta = 1-\sqrt{\log n /T}$ as in Ref.~\cite{klivans2017learning}.
\begin{figure}[h!]
\begin{algorithm}[H]
  \caption{\textsc{Sparsitron}  $\left(T,M, \lambda, \left(X^{(i)},Y^{(i)}\right)_{i=1}^{T+M}\right)$ } \label{algosparsitron}
  \begin{algorithmic}[1]
    \Require{$T+M$ independent samples $( X^{(i)},Y^{(i)})$ for $i\in [T+M]$, $ \lambda >0$.}
    \State $\beta \gets 1-\sqrt{\log n /T}$.
    \State $w^{(0)}\gets \textbf{1}/n$.
    \For{$t =1,...,T$}
    \State $\alpha^{(t)} \gets w^{(t)}/\Vert w^{(t)} \Vert_1$. 
    \State   $l^{(t)}  \gets 1/2 \left(\textbf{1} + \left(\sigma ( \lambda \alpha^{(t)} \cdot X^{(t)} )-Y^{(t)} \right)X^{(t)} \right)$.
    \State For each $i\in[n]$,  $w_i^{(t)}  \gets w_i^{(t-1)} \cdot \beta^{l^{(t)}_i}$.
    \EndFor
    \For{$t =1,...,T$}
    \State Compute the empirical risk 
    $$ \varepsilon \left(\lambda \alpha^{(t)} \right)\gets \frac{1}{M}\sum_{m=1}^{M}\left( \sigma \left( \lambda \alpha^{(t)} \cdot X^{(T+m)}\right)-Y^{(T+m)} \right)^2 .$$
    \EndFor 
    \Ensure {$v=\lambda \alpha^{(t')}$ for $t'=\text{argmin}_{t\in[T]} \varepsilon \left(\lambda \alpha^{(t)} \right)$.}
  \end{algorithmic}
\end{algorithm}
\end{figure}

The following lemmas have been proved in \cite{klivans2017learning}. 

\begin{lemma}[\cite{klivans2017learning}]\label{lemma_MRF_function_unbiased}
Let $\cal{D}$ be a $r$-wise MRF on $\{-1,1\}^n$ with underlying dependency graph $G$ and factorization polynomial $p(z) = \sum_{I \in C_t(G)} p_I(z)$ with $\max_i \|\partial_i p\|_1 \leq \lambda$. Then, the following hold for $Z \sim \cal{D}$:
\begin{itemize}
\item For any $i$, and a partial assignment $x \in \{1,-1\}^{[n] \setminus \{i\}}$, $Pr[Z_i = -1 | Z_{-i} = z] = \sigma(- 2 \partial_i p(z))$, where $\sigma(z)$ is the sigmoid function. 
\item $\cal{D}$ is $e^{-2\lambda}/2$-unbiased.
\end{itemize}
\end{lemma}

\begin{lemma}
[Lemma \uppercase\expandafter{\romannumeral 6}.3 \cite{klivans2017learning}]\label{lemma_6.3}
Let $\cal{D}$ be a $\delta$-unbiased distribution on $\{1,-1\}^n$.  Let $p,q$ to be two multi-linear polynomial  $p,q : \mathbb{R}^n \to \mathbb{R}$ such that   $\mathbb{E}_{X\sim \cal{D}} \left[\left(\sigma( p(x)) -\sigma(q(x))\right)^2\right]\leq \epsilon$ where $\epsilon< e^{-2\Vert p \Vert_1 -6}\delta^{\vert I\vert}$. Then for every maximal monomial $I \subseteq [n]\setminus \{u\}$ of $p-q$,
\begin{eqnarray}
\vert \widehat{p}(I)-\widehat{q}(I)\vert \leq  e^{\Vert p \Vert_1 +3}\sqrt{\epsilon / \delta^{\vert I\vert}}.
\end{eqnarray}
\end{lemma}

\section{Quantum Sparsitron Algorithm}
\label{apendix_quantum_sparsitron}
The quantum version of the \textsc{Sparsitron} algorithm is shown in Algorithm \ref{algSparsitronQuantum}.
 \begin{center}
\begin{figure}[!h]
\begin{minipage}{\linewidth}
\begin{algorithm}[H]
  \caption{\textsc{Quantum\_Sparsitron}  $\left(T,M, \epsilon,  \delta, \lambda, \left(x^{(i)}, y^{(i)}\right)_{i=1}^{T+M} \right)$ }
    \label{algSparsitronQuantum}
  \begin{algorithmic}[1]
    \Require{ Error $\epsilon\in (0,1)$, a function $\sigma$: $\mathbb R\rightarrow [0,1]$,  probability $\delta \in(0,1)$, norm $\lambda\geq 0$, quantum access to training set $(x^{(i)}, y^{(i)}) \in [-1,1]^N \times \{0,1\}$ for $i\in[T+M].$
     }
     $\beta \gets 1-\sqrt{\log n /T}$
    \For{$t = 1 \textrm{ to } T$}
    \State Construct unitaries for $w^{(t)} = w^{(1)} \beta^{\widetilde l^{(1)}}\cdots \beta^{\widetilde l^{(t-1)}}$ using unitaries $\{U^{(t')}_{\rm loss} : t' \in [t-1] \}$.
     \State $ h^{(t)} \gets$  Estimate $\frac{w^{(t)}}{ \Vert w^{(t)} \Vert_1} \cdot x^{(t)}$ to additive accuracy $ \frac{\epsilon}{8\lambda^2}$ with success probability $1-\frac{\delta}{2T}$.
      \For{$m=1 \textrm{ to } M$} 
     \State $z^{(t,m)} \gets$ Estimate $\frac{ w^{(t)} }{\Vert w^{(t)}\Vert_1} \cdot x^{(T+m)}$ to additive accuracy $\frac{\epsilon}{16 \lambda}$ with success probability $1-\frac{\delta}{2MT}$.
    \EndFor
    \State $\widetilde \varepsilon^{(t)} \gets \frac{1}{M} \sum_{m=1}^{M}\left(\sigma \left(\lambda  {z^{(t,m)}}\right ) - y^{(T+m)}\right)^2$.
    \State Construct unitary $U^{(t)}_{\rm loss}$ that prepares $\ket j \ket {\widetilde l^{(t)}_j}$ with $\widetilde l^{(t)} = \frac{1}{2}\left ( \vec 1 +\left(\sigma\left (\lambda { h^{(t)}}\right ) - y^{(t)}\right) x^{(t)}\right)$ for the next step.
    \EndFor
    \State $t' =\arg \min_{t\in[T]}\widetilde \varepsilon^{(t)}$.
    \State $\Gamma^{(t')} \gets$   Estimate of $\Vert w^{(t')} \Vert_1 $ by determining $w_{\max}^{(t')} $  and estimating $ \left\Vert \frac{ w^{(t')}}{ w_{\max}^{(t')} } \right \Vert_1$. 
     \Ensure {$\left (t', {h^{(1)}}   \cdots, {h^{(t')}} , \Gamma^{(t')} \right)$. }
  \end{algorithmic}
\end{algorithm}
\end{minipage}
\end{figure}
\end{center}

\section{Quantum subroutines}

\begin{lemma}[Quantum counting \cite{brassard2002quantum}]
\label{lemma_quantum_counting}
Given a Boolean function $f:\{0,1,...,N-1\}\rightarrow \{0,1\}$,  $t=|f^{-1}(1)|$,   there is an algorithm which outputs $t'$ which equal to $t$ 
with probability $1-\rho$ with run time $\Ord{\sqrt{tN}\log \frac{1}{\rho}}$. If $t=0$, we can obtain $t'=t$ with certainty with run time $\Ord{\sqrt{N}}$. 
\end{lemma}

 \begin{lemma}[Quantum search \cite{boyer1998tight}]\label{Tight_bound_quanutm_search}
Given quantum access to $N$ numbers, if there are $M$ solutions of a search problem, there is a quantum algorithm  finds a solution in time $\Ord{\sqrt{\frac{M}{N}}\log\frac{1}{\rho}}$with success probability $1-\rho$.
\end{lemma}

\section{Quantum set membership}
\label{appendix_quantum_indicator_function}
We first give the definition of QRAM for a vector in the following. 
\begin{defn}[Quantum RAM \cite{giovannetti2008quantum,giovannetti2008architectures,arunachalam2015robustness}]
	\label{defQRAM}
Let $c$ and $n$  be positive integers.
Let us be given a vector $v$ of dimension $n$,
where each element of $v$ is a bit string of length $c$, i.e., $v \in \mathbbm (\{0,1\}^{c})^{n}$. Quantum RAM takes $v$ as input with a one-time cost of $\tOrd{c n}$. Then, with an arbitrary bit string $b\in\{0,1\}^c$, quantum RAM provides the operation
\be
\ket {j} \ket{b} \to \ket{j} \ket {b \oplus v_j}, 
\ee
at a cost of $\Ord{c+\log^2 n}$.
\end{defn}
Note that calling this operation twice leads to $\ket {j} \ket{b} \to \ket {j} \ket{b}$.

\begin{lemma} \label{QRAM_construction}
Assume the availability of a Quantum RAM as in Definition \ref{defQRAM}. Given a set $S \subset [N]$ with size $|S|=\mu$, let $S_1\leq S_2 \cdots \leq S_\mu$ be the sorted elements of the set. We can construct a sorted QRAM for set $S$ in time $\tOrd{\mu \log N}$. With an arbitrary bit string $b\in\{0,1\}^{\lceil \log N \rceil}$, this QRAM provides the operation
\be
\ket {k} \ket{b} \to \ket{k} \ket {b \oplus S_k}, 
\ee
where $k \in [K]$ with $K = 2^{\lceil \log \mu \rceil}$ and with the zero-padding $S_{\mu+1}, \cdots, S_K = 0$, at a cost of 
$\Ord{\log N + \log^2 \mu}$ per query. 
\end{lemma}
\begin{proof}
Let $m:=\lceil \log \mu \rceil,$  we extend the set $S$ by dummy elements $0$ such that its size is $2^m$.
Store the extended $S$ into QRAM from Definition \ref{defQRAM}.
Using merge sort costs  $\Ord{\mu \log \mu}$ time and space.
The construction of  a QRAM with entries of size $\Ord{\log N}$ for $m \equiv \lceil \log \mu \rceil$ register qubits requires time and space of $\tOrd{2^m \log N}=\tOrd{\mu \log N}$.
\end{proof}
Using this Lemma, we can provide our result on quantum set membership queries. 
For the purpose of this work, it is sufficient to use for the run time per query the bound
$\subseteq \Ord{\log^3 N}$.
\begin{theorem}[Quantum set membership] \label{Quantum_indicator_function}
Given a QRAM for set $S \subset [N]$ with size $|S|=\mu $ according to Lemma \ref{QRAM_construction}, we can provide a quantum unitary $U_{S}$ which performs the quantum indicator function (or set membership query)
\begin{eqnarray}
 \ket{j}\ket{0}_{\rm result}\to
\begin{cases}
 \ket{j}\ket{1}_{\rm result}, &\text{for} ~ j\in S\\
 \ket{j}\ket{0}_{\rm result}, & \text{for} ~ j\notin S
\end{cases}
\end{eqnarray}
with time $\Ord{\log\mu\log N+\log^3\mu }$.  
\end{theorem}
\begin{proof}

We can query access the $k$-th element of $S$ with the operation
\begin{eqnarray}
\ket{k}\ket{\bar{0}}\to \ket{k}\ket{S_k}.
\end{eqnarray}
The method is inspired by  binary search. 
For binary search, define the comparison operation for all $j,j' \in[N]$,
\be
T(j, j') := \begin{cases} 0 & {\rm if}\ j\leq j' \\ 1 & {\rm if }\ j>j' \end{cases}.
\ee
The binary search involves a pivot element which is compared to the input element. 
We take a qubit register $\ket{\bar 0}_{\rm A}$ to represent the address of this pivot in the QRAM and 
the pivot itself in register $\ket{\bar 0}_{\rm B}$.
The starting point for the binary search is the address state $\ket{01\cdots 1}_{\rm A}$ which refers to the $2^{m-1}$-th  element of $S$ (about the ``middle" of the array).
Querying the QRAM with this state obtains $ \ket{01\cdots 1}_{\rm A} \ket{S_{2^{m-1}}}_B$.
With the input state, perform the comparison ($j\in[N]$) in another qubit (denoted by $C$)
\be
 \ket j \ket{01\cdots 1}_{\rm A} \ket{S_{2^{m-1}}}_B \ket{T(j, S_{2^{m-1}})}_C.
\ee
Undoing the QRAM leads to
\be
 \ket j \ket{01\cdots 1}_{\rm A} \ket{\bar 0}_B \ket{T(j, S_{2^{m-1}})}_C.
\ee
Now perform a SWAP operation with the qubit $C$ and 
the first qubit of register $A$
\be
 \ket j \ket{T(j, S_{2^{m-1}}),1\cdots 1}_{\rm A} \ket{\bar 0}_B \ket{0}_C.
\ee
We have an updated $A$ register state and, as in binary search, we want to compare with the 
$2^{m-2}$-th element and the
$2^{m-1}+2^{m-2}$-th element of the set in superposition. Hence we flip the second bit in the address register to $0$ with an $X$ gate to obtain
\be
 \ket j \ket{T(j, S_{2^{m-1}}),0, 1\cdots 1}_{\rm A} \ket{\bar 0}_B \ket{0}_C.
\ee
Depending on $T(j, S_{2^{m-1}})$, this address corresponds precisely to the elements at position $2^{m-2}$ and
$2^{m-1}+2^{m-2}$.
We now query the QRAM at this address and perform the comparison again. Recurse for $m$ steps.
This achieves
\be
 \ket j \ket{T_{j,1},T_{j,2},\cdots, T_{j,m}}_{\rm A} \ket{\bar 0}_B \ket{0}_C.
\ee
Now perform another query of the QRAM with the address $\ket{T_{j,1},T_{j,2},\cdots, T_{j,m}}_{\rm A}$, to obtain the element $j^\ast$ in the register $B$. Then, compare $j$ in the first register and $j^\ast$ in register $B$, and output in the result register as
\be
 \ket j \ket{T_{j,1},T_{j,2},\cdots, T_{j,m}}_{\rm A} \ket{j^\ast}_B \ket 0_C \ket{ j =j^\ast}_{\rm result}.
\ee 
We repeat the discussion of same steps of this procedure in a bit more detail.
Let an arbitrary quantum state be given as $\sum_{j = 1}^N \alpha_j\ket{j}$ with amplitudes $\alpha_j \in \mathbbm C$, where $\Vert \alpha \Vert_2 = 1$.   
\begin{itemize}
\item[(1)]  Prepare an $m$-qubit register $A$ with all qubits in state $\ket{1}$, an $\Ord{\log N}$-qubit register $B$ in state $\bar{0}$, and a single qubit $C$ in state $\ket{0}$. The state is given by:
\begin{eqnarray}
\sum_{j =1}^N\alpha_j\ket{j}\ket{1\cdots 1}_{A} \ket{\bar{0}}_{B}\ket{0}_{C} \ket{0}_{\rm result}.
\end{eqnarray}
\item[(2)] For $t=1$ to $m$,  do: 
\begin{itemize}
\item[\romannumeral1.] Apply a $X$ gate on the  $t$-th qubit of register  $A$.
\item[\romannumeral2.]  Query access to $S$ with address register $A$ and value register $B$. 
\item[\romannumeral3.] Compare the value $j$ with the value in register $B$, denoted by  $j'$. If $ j > j'$, apply a $X$ gate on the qubit $C$.
 \item[\romannumeral4.] Undo the oracle and swap the $t$-th qubit in register $A$ with the qubit $C$.
\end{itemize}
\item[(3)]  Query access to $S$ with register $A$ and $B$ to obtain value $j^\ast$ in register B. Quantum compare the value $j$ with $j^\ast$.  
If $j =j^\ast$, apply a $X$ gate on the qubit  $\ket {0}_{\rm result}$. Undo the query.
\item[(4)] This step is for undoing the Step 2. For $t=m$ to $1$,  do 
\begin{itemize} 
\item[\uppercase\expandafter{\romannumeral1}.] Swap the $t$-th qubit in register $A$ with the qubit $C$.
\item[\uppercase\expandafter{\romannumeral2}]  Query access to $S$ with register $A$ and $B$. 
\item[\uppercase\expandafter{\romannumeral3}] Compare the value $j$ with the value in register $B$ called $j'$. If $ j > j'$, apply a $X$ gate on the ancillary qubit  $C$.
 \item[\uppercase\expandafter{\romannumeral4}]  Undo the query and apply a $X$ gate on the  $t$-th qubit of the register  $A$.
\end{itemize}
\end{itemize}
These steps obtain 
\begin{eqnarray}
\sum_{j \in S} \alpha_j \ket{j}\ket{1}_{\rm result}+\sum_{j\not \in S } \alpha_j\ket{j}\ket{0}_{\rm result}.
\end{eqnarray}

We now analyse the run time for this implementation of the set membership query. For the quantum membership query, the step $(1)$ takes time $m$ to obtain 
 $\ket{1\cdots 1}_{A} $ via $X$ gates. Each $t \in [m]$ of the step $(2)$ requires $\Ord{\log N + m^2}$ time as step \romannumeral1, \romannumeral2, \romannumeral3, \romannumeral4,  takes $1$, $\Ord{\log N + m^2}$, $\Ord{\log N}$, $\Ord{m}$ gates respectively, hence $\Ord{\log N + m^2}$ in total. Step $(2)$ in total requires hence $\Ord{m(\log N +m^2)}$ quantum gates. Step $(3)$ costs $\Ord{\log N + m^2}$ run time for the query and the comparison. Step $(4)$ is the reverse of Step $(2)$ and hence requires $\Ord{m(\log N +m^2)}$  time.  In total, the run time for performing the unitary $U_{S}$ is $\Ord{\log \mu(\log N +\log^2 \mu)}.$
\end{proof}

For example given $S=\{001,011,101\} = \{1,3,5 \}$ and a state $\sum_{j=1}^7\ket{j}\ket{0}_{\rm result}$, we first extend $S$  to $S=\{001,011,101, 000\}$ such that $|S|=2^2$, then we construct  a QRAM for $S$. e have the following states by using above procedure. First we prepare state  $\sum_{j=1}^7\ket{j}\ket{11}_{A} \ket{000}_{B}\ket{0}_{C} $ by using the above procedure,  for $t=1$, we run step  \romannumeral1, \romannumeral2, \romannumeral3, \romannumeral4, for simplification, we ignore the normalization factor $\frac{1}{\sqrt{7}}$
\begin{eqnarray}
 \sum_{j=1}^7\ket{j} \ket{11}_{A} \ket{000}_{B}\ket{0}_{C} 
&\xrightarrow{~ \text{\romannumeral1~}} &  \sum_{j=1}^7\ket{j} \ket{01}_{A} \ket{000}_{B}\ket{0}_{C}\nonumber\\ 
& \xrightarrow{~ \text{\romannumeral2~}} & \sum_{j=1}^7\ket{j}\ket{01}_{A} \ket{S_2=011}_{B}\ket{0}_{C}\nonumber\\ 
& \xrightarrow{~ \text{\romannumeral3~}}& \sum_{j\leq 3}^7\ket{j}\ket{01}_{A} \ket{011}_{B}\ket{0}_{C}+\sum_{j>3}^7\ket{j}\ket{01}_{A} \ket{011}_{B}\ket{1}_{C}\nonumber\\ 
& \xrightarrow{~ \text{\romannumeral4~}}&  \sum_{j\leq 3}^7\ket{j} \ket{01}_{A} \ket{000}_{B}\ket{0}_{C}+\sum_{j>3}^7\ket{j}\ket{11}_{A} \ket{000}_{B}\ket{0}_{C}.
\end{eqnarray}

Then run  step  \romannumeral1, \romannumeral2, \romannumeral3, \romannumeral4,   for $t=2$
\begin{eqnarray}
& \xrightarrow{~ \text{\romannumeral1 ~} }&   \sum_{j\leq 3}^7\ket{j}\ket{00}_{A} \ket{000}_{B}\ket{0}_{C}+\sum_{j>3}^7\ket{j}\ket{10}_{A} \ket{000}_{B}\ket{0}_{C} \nonumber\\ 
&\xrightarrow{~ \text{\romannumeral2~}}  & \sum_{j\leq 3}^7\ket{j}\ket{00}_{A} \ket{S_1=001}_{B}\ket{0}_{C}+\sum_{j>3}^7\ket{j}\ket{10}_{A} \ket{S_3=101}_{B}\ket{0}_{C}\nonumber\\ 
& \xrightarrow{~ \text{\romannumeral3~}}   & \sum_{j\leq 1}^7\ket{j}\ket{00}_{A} \ket{001}_{B}\ket{0}_{C}
 + \sum_{1<j\leq 3}^7\ket{j}\ket{00}_{A} \ket{001}_{B}\ket{1}_{C}\nonumber\\
& &+\sum_{3<j \leq 5 }^7\ket{j}\ket{10}_{A} \ket{101}_{B}\ket{0}_{C}
+\sum_{j>5}^7\ket{j}\ket{10}_{A} \ket{101}_{B}\ket{1}_{C}  \nonumber\\
& \xrightarrow{~\text{\romannumeral4~} }  & \sum_{j\leq 1}^7\ket{j}\ket{00}_{A} \ket{000}_{B}\ket{0}_{C}
+ \sum_{1<j\leq 3}^7\ket{j}\ket{01}_{A} \ket{000}_{B}\ket{0}_{C}\nonumber\\
& &+\sum_{3<j \leq 5 }^7\ket{j}\ket{10}_{A} \ket{000}_{B}\ket{0}_{C}
+\sum_{j>5}^7\ket{j}\ket{11}_{A} \ket{000}_{B}\ket{0}_{C}
\end{eqnarray}

Query access to $S$ with register $A$ and $B$ as step (3) of the procedure,  we have 
  \begin{eqnarray}
\sum_{j\leq 1}^7\ket{j}\ket{00}_{A} \ket{001}_{B}\ket{0}_{C}\ket{0}_{result}
+ \sum_{1<j\leq 3}^7\ket{j}\ket{01}_{A} \ket{011}_{B}\ket{0}_{C}\ket{0}_{result}\nonumber\\
+\sum_{3<j \leq 5 }^7\ket{j}\ket{10}_{A} \ket{101}_{B}\ket{0}_{C}\ket{0}_{result}
+\sum_{j>5}^7\ket{j}\ket{11}_{A} \ket{000}_{B}\ket{0}_{C}\ket{0}_{result}
\end{eqnarray}
Compare the value $j$  with the value $j^*$  in register $B$, if $ j =j^*$, the qubit $\ket{0}_{\rm result}$ turn to state $\ket{1}_{\rm result} $,  undo the query, it yields
  \begin{eqnarray}
  & &\left( \ket{001}\ket{00}_{A}+\ket{011}\ket{01}_{A} +\ket{101}\ket{10}_{A}\right) \ket{000}_{B}\ket{0}_{C}\ket{1}_{\rm result}\nonumber\\
  &+&  \left(\ket{010} \ket{01}_{A}+\ket{100}\ket{10}_{A} + \ket{110}\ket{11}_{A}+\ket{111} \ket{11}_{A}\right)  \ket{000}_{B}\ket{0}_{C}\ket{0}_{\rm result}
\end{eqnarray}

Performing step $(4)$,  we obtain
\begin{eqnarray}
\sum_{j\in S}\ket{j}\ket{1}_{\rm result}+\sum_{j\notin S}\ket{j}\ket{0}_{\rm result}.
\end{eqnarray}

\end{document}